\documentclass[
reprint,
superscriptaddress,
amsmath,
amssymb,
aps,
prb,
floatfix
]{revtex4-2}
\usepackage{graphicx}
\usepackage{hyperref}
\usepackage{xcolor}
\usepackage{multirow}
\usepackage{enumitem}

\begin{document}

\title{Optimal parallelisation strategies for flat histogram Monte Carlo sampling}

\author{Hubert J. Naguszewski}
\email[Corresponding Author\\]{Hubert.Naguszewski@warwick.ac.uk}
\affiliation{Department of Physics, University of Warwick, Coventry, CV4 7AL, United Kingdom}
\author{Christopher D. Woodgate}
\email{christopher.woodgate@bristol.ac.uk}
\affiliation{H.H. Wills Physics Laboratory, University of Bristol, Royal Fort, Bristol, BS8 1TL, United Kingdom}
\author{David Quigley}
\email[]{D.Quigley@warwick.ac.uk}
\affiliation{Department of Physics, University of Warwick, Coventry, CV4 7AL, United Kingdom}

\begin{abstract}
Flat histogram methods, such as Wang--Landau sampling, provide a means for high-throughput calculation of phase diagrams of atomistic/lattice model systems. Many parallelisation schemes with varying degrees of complexity have been proposed to accelerate such sampling simulations. In this study, several widely used schemes are benchmarked---both in isolation and in combination---to establish best practice. The schemes studied include energy domain decomposition with both static sizing of energy sub-domains, as well as a dynamic sub-domain sizing scheme which we propose. We also assess the benefits both of replica exchange and of including multiple random walkers per sub-domain, to determine which factors have the largest impact on parallel efficiency. Additionally, the influence of energy sub-domain overlap regions is discussed. As illustrative test cases, we implement and apply the aforementioned strategies to a lattice-based model describing the internal energy of a substitutional alloy, studying the AlTiCrMo refractory high-entropy superalloy as well as the binary CuZn system, both of which crystallographically order into a B2 (CsCl) structure with decreasing temperature. We find that---while all of the proposed strategies confer a non-negligible speedup---parallelisation across energy domains which are non-uniform in size offers the most appreciable performance improvements. This work offers concrete recommendations for which parallelisation strategies should be prioritised to optimally accelerate flat-histogram Monte Carlo simulations.
\end{abstract}

\date{February 9, 2026}

\maketitle


\section{Introduction}

Monte Carlo (MC) methods are the workhorse tool for studying the phase diagrams, thermodynamics and phase equilibria of a wide range of atomistic/lattice models of materials~\cite{landau_guide_2021}. However, a longstanding limitation of standard MC methods, based on the Metropolis algorithm~\cite{metropolis_equation_1953}, is their restriction to obtaining results for only a single simulation temperature for a given simulation instance, and their inefficiency (particularly their slow rate of convergence) near discontinuous phase transitions. Near such discontinuous transitions, it is typically found that accurate characterisation of sharp features in state functions, such as peaks in specific heat, often requires a large number of independent simulations, each of which must utilise a substantial number of trial MC moves. To overcome this issue, several so-called `enhanced' sampling techniques have been developed, including umbrella sampling, transition matrix MC, and Wang--Landau sampling~\cite{frenkel_understanding_2002, bennett_efficient_1976, torrie_nonphysical_1977, berg_multicanonical_1991, berg_multicanonical_1992, lee_new_1993, wang_transition_1999}.

The root cause of the inefficiency of conventional sampling techniques in some settings is the fact that, in complex materials, ergodicity can be severely limited. This is a particularly significant problem when large free energy barriers separate regions of configuration space. Umbrella sampling (for example) addresses this by introducing a bias potential that facilitates transitions across these barriers, enabling more uniform sampling throughout the configuration space of the model. This method is part of a broader class of re-weighting techniques, in which the usual Boltzmann factor is replaced by a modified weight function, in order to remove the `real' thermodynamics such that a more uniform sampling is obtained. From this (biased) more uniform sampling, the (unbiased) Boltzmann statistics are subsequently recovered during post-processing.

Building on this concept, the Wang--Landau (WL)~\cite{wang_efficient_2001, wang_determining_2001, landau_new_2004, landau_determining_2002, brown_improved_2011} sampling algorithm can be used to obtain the density of states (DOS) of a given model in the canonical ensemble, which can then be used to compute physical properties such as specific heat and internal energy---both as functions of temperature---as a simple post-processing step. The WL algorithm samples configurations in inverse proportion to the DOS, an initial estimate of which is iteratively refined during the sampling process until a uniform sampling distribution in energy is obtained. In this way, from a single simulation, one obtains both a set of samples which span the entire energy domain of interest, and the DOS over that same energy domain. WL sampling has found utility across a range of research areas, including the study of phase transitions~\cite{volkov_phase_2012, lee_frustrated_2024, sato_wanglandau_2010, pei_error_2019}, polymer physics~\cite{wang_phase_2011, antypov_computer_2008, parsons_off-lattice_2006}, solubility prediction~\cite{boothroyd_solubility_2018, boothroyd_solubility_2019}, magnetic materials~\cite{bin-omran_influence_2017, nguyen_study_2017, miyashita_atomistic_2021, bin-omran_wang-landau_2016, ngo_phase_2008, bogoslovskiy_phase_2024}, and surface adsorption (adhesion of atoms/ions/molecules to surfaces)~\cite{allen_wang-landau_2012, lazo_phase_2009, swetnam_improved_2009}. Methods to update the statistical temperature of the system instead of the DOS within WL have also been proposed and used to develop an algorithm equivalent of WL for molecular dynamics known as statistical temperature molecular dynamics (STMD)~\cite{kim_statistical-temperature_2006}. 

Various schemes have been proposed to parallelise the WL algorithm~\cite{wl_mpi_2013}, and more broadly to parallelise other classes of multicanonical simulations~\cite{zierenberg2013, gross2018}, with the aim of optimising utilisation of compute resource and minimising simulation runtime. However, there is no consensus on which scheme---or combination of schemes---is the most efficient and offers the most speedup per WL instance, where each instance corresponds to one CPU core in the context of the present work. (This is because evaluation of our model Hamiltonian is performed using a single CPU core.) A common feature among many proposed parallelisation strategies is the decomposition of the energy domain (the total range of energies accessible to the simulation system) into energy sub-domains (smaller windows of energy). The WL algorithm is then applied concurrently in each energy sub-domain, and subsequently these are combined to produce the DOS across the entire energy range of interest.

This paper aims to identify the most efficient parallel implementation of the WL algorithm, with a particular focus on iterative runtime load balancing. The context of our work is the need to perform a great many high-throughput WL calculations with varying Hamiltonian parameters and material compositions. We are hence primarily concerned with maximising per-core parallel efficiency rather than reaching the highest possible speedup or core count for a single WL calculation. A key development of the present work is the demonstration that, despite the challenging nature of achieving perfectly optimal load balancing, it can be closely approached by dynamically adjusting the size of each energy sub-domain after each WL iteration, thereby leading to improved computational performance over long simulation runs. We also demonstrate how the performance of the algorithm can dramatically depend on several factors, including: the choice of the number of energy sub-domains and their size; the number of walkers within each energy sub-domain; and the size of the overlap between neighbouring energy sub-domains.

The remainder of this manuscript is structured as follows. First, in Sec.~\ref{sec:wl_sampling} we give an overview of the WL sampling algorithm. Subsequently, in Sec.~\ref{sec:parallel_implementation}, we discuss the various means by which the algorithm may be parallelised. Following on from this, in Sec.~\ref{sec:bw_model_implementation} we present the two model systems used for benchmarking---a high-entropy alloy and a binary alloy---and how they were implemented, along with a brief reprise of metrics for assessing parallel efficiency with respect to number of sub-domains and walkers. Proceeding in Sec.~\ref{sec:results}, we present results comparing the efficiency of parallelisation schemes in the context of a benchmark model system, both in isolation and in combination. Finally, in Sec.~\ref{sec:conclusions_and_recommendations}, we conclude with recommendations on which parallelisation scheme(s) to prioritise when implementing flat histogram MC methods. Throughout the paper we will note where we expect conclusions to be specific to the system and where they are more generally applicable.

\section{Methods}

\subsection{Wang--Landau (WL) Sampling}
\label{sec:wl_sampling}

WL sampling~\cite{wang_efficient_2001, landau_guide_2021} is a flat energy histogram technique with broad applicability. For a given model system---with an associated Hamiltonian that defines the energy as a function of the model's configuration---the algorithm determines the DOS, from which thermodynamic quantities can subsequently be calculated for any temperature. This is in contrast to the simpler Metropolis MC algorithm, in which a separate simulation is required at every discrete temperature of interest. In order to obtain the DOS, we must first define an accessible energy domain of interest. This is perhaps most easily found by performing a simulation using the Metropolis algorithm to determine the extrema of the energy at the highest ($T_{\text{max}}$) and lowest ($T_{\text{min}}$) temperatures of interest. Note that the lowest temperature of interest is not necessarily one at which the ground state of the model system has non-vanishing statistical weight if, for example, the simulation is being performed to locate phase transitions at higher temperatures. Using the energies found as a guide, we can choose an energy domain which is uniformly divided into $N$ bins of equal size, with each bin representing an energy macrostate containing multiple model system configurations. Given a model in which microstates are discrete---such as the lattice-based models in this study---the partition function $Z$ is expressed as a sum over all microstates. This can be re-written as a sum over all the discrete energies $E$~\cite{landau_guide_2021}, \textit{i.e.}
\begin{equation}
    Z = \sum_i \exp(-E_i / k_B T)
    \equiv \sum_E g(E)\, \exp(-E / k_B T),
    \label{eq:microstate_sum}
\end{equation}
where $E_i$ is the energy of the $i^{\text{th}}$ microstate, $k_B$ is the Boltzmann constant, $T$ is the temperature, and $g(E)$ is the density of states corresponding to energy $E$. The expression on the right of Eq.~\ref{eq:microstate_sum} is then approximated and re-written as a sum over `binned' energy macrostates, taking the form
\begin{equation}
    Z \approx \sum_j g(E_j)\, \exp(-E_j / k_B T),
    \label{eq:macrostate_sum}
\end{equation}
where $E_j$ is the energy of the centre of the $j^{\text{th}}$ energy bin, and $g(E_j)$ is the DOS corresponding to the  $j^{\text{th}}$ energy bin. (Note that the first summation appearing in Eq.~\ref{eq:microstate_sum} is the usual summation over microstates, and that the second summation of Eq.~\ref{eq:microstate_sum} is taken over \textit{all} energies accessible to the system. This is in contrast to the approximate summation of Eq.~\ref{eq:macrostate_sum}, which is a sum over the newly-defined energy bins, taken to represent macrostates.) If the system being studied is particularly discrete, then it is possible for energy states accessible to the system to be spaced further apart than the width of the WL energy bins. This can result in energy bins that have no energy states and are therefore not possible to visit, which would interfere with WL sampling. To remedy this, one can simply adjust the binning such that there are no bins without available energy states. In general, we expect this to not be an issue for systems that will benefit from the parallelisation schemes outlined within this study.

In order to obtain the DOS, the WL algorithm performs a random walk in the configuration space of the model. Trial moves from an initial configuration, labelled $n$, to a proposed configuration, labelled $m$, are accepted with probability
\begin{equation}
    P_{n\rightarrow m} = 
    \begin{cases}
        \; \frac{g(E_n)}{g(E_m)}, &\quad g(E_n) < g(E_m)\\
        \; 1, &\quad g(E_n) \geq g(E_m),
    \end{cases}
\label{eq:wl_transition_probability}
\end{equation}
where $E_n$ is the energy of the initial configuration and $E_m$ is the energy of the proposed configuration. In the context of the fixed-lattice high-entropy alloy simulations considered in this work, MC trial moves consist of two randomly selected atoms within the system swapping positions, accepted according to the probability in Eq.~\ref{eq:wl_transition_probability}. This algorithm is not self-starting however, as the DOS is not known at the start of the simulation and an initial estimate for the DOS is required. This is usually taken as $g(E) \equiv 1$, and is iteratively refined throughout the simulation, with the uniformity of sampling being a measure of how accurately the DOS is currently estimated. Once the DOS is sufficiently accurate, a uniform distribution of samples is obtained. The estimate of the DOS is refined in iterations by modifying the DOS after each trial move according to
\begin{equation}
    g(E_i) \rightarrow g(E_i)f_k,
\end{equation}
with $E_i$ being the energy of the resulting state, while $f_k$ is known as the modification factor at WL iteration $k$, which in practice takes a value close to, but slightly greater than, unity. (Note that this means that there are no `wasted' moves; even a rejected move results in the DOS being updated.) Due to the many orders of magnitude over which the DOS varies, it is often numerically convenient to work with $\log{g(E)}$ rather than $g{(E)}$ itself, incremented by $\log(f_k)$ at each trial move. To further address the issue of magnitude, the logarithm of the DOS is periodically shifted such that the minimum value is zero. At the start of the simulation, the iteration number is $k=1$ and $\log (f_1)$ is set to some appropriately small value---we choose $\log(f_1)=0.05$. A histogram of the visited energies, $H(E)$, is maintained, and the WL iteration number is incremented once the histogram reaches a desired level of `flatness', $F$, defined as
\begin{equation}
    F = \frac{\min(H(E))}{\frac{1}{N}\sum_i^N H(E_i)},
\end{equation}
where the target value of $F$ is typically chosen to be close to 0.8 (80\%), indicating that the simulation is now exploring all energy windows near-equally. If this flatness condition is met, the sampling is paused, and the modification factor $f$ is reduced to be closer to unity for the next iteration, \textit{e.g.} $f_{k+1}=\sqrt{f_k}$. The histogram, $H(E)$, tracking the visited energies is reset to be uniformly zero, and the sampling continues until $f$ falls below a set tolerance $f = 1 + \varepsilon$ for some appropriately small value of $\varepsilon$, defining the degree of convergence of the DOS. The DOS is then taken to be that obtained after the final WL iteration. In this work, we take $\varepsilon = 2\cdot 10^{-5}$ purely as a convenient level of convergence to use as a benchmark. In practice, $\varepsilon$ could be taken as much smaller than this for generating accurate results, and convergence of any calculated output quantities for a given model system should always be thoroughly tested.

Once the DOS for the system is obtained, the energy probability distribution at a particular temperature can be calculated as a simple post-processing step,
\begin{equation}
    P(E_i, T) = \frac{g(E_i) \exp(-E_i/k_BT)}{Z},
    \label{eq:prob_dist}
\end{equation}
where $Z$ is calculated according to Eq.~\ref{eq:macrostate_sum}. From this distribution, various system properties can then be derived. An example of such a property is the specific heat, given by
\begin{equation}
    C = \frac{\langle E^2\rangle - \langle E\rangle^2}{k_BT^2},
\end{equation}
where $\langle E^2 \rangle$ and $\langle E \rangle$ are the average values of the square of the simulation energy and the simulation energy at a particular temperature, respectively. These averages can be calculated using the energy probability distribution of Eq.~\ref{eq:prob_dist}. In this manner, it can be seen that, once a sufficiently accurate estimate of $g(E)$ has been obtained, it is possible to run a simulation which samples all energies and temperatures of interest without further modification of the DOS ($f=1$). In principle, the distribution for any observable sampled during this phase of the simulation (\textit{e.g.}, a choice of order parameter describing a phase transition) can also be re-weighted to calculate its value at any temperature of interest.

Independently of any parallelisation strategy, there have been efforts to optimise both the algorithm's `modification factor' and `flatness criterion'~\cite{zhou_understanding_2005, belardinelli_fast_2007, zhou_wang-landau_2006, swetnam_improving_2011}. The WL algorithm was quickly improved by the introduction of energy domain decomposition into sub-domains~\cite{wang_efficient_2001}. This decomposition allowed for independent random walks to be performed either in serial or concurrently. Further optimisation can be achieved by dynamically adjusting the size of each energy sub-domain; however, achieving maximally optimal load balancing is challenging due to the complex nature of the free energy landscape for a given system, and several methods have been proposed for the dynamic adjustment of sub-domain size within a WL simulation~\cite{cunha-netto_critical_2011, cunha-netto_improving_2008, cunha-netto_two-dimensional_2009}. These methods typically rely on restricting the energy domain over which a single WL random walk occurs. Once the flatness criterion has been met on a subset of the total energy domain, that portion of the energy domain is excluded from the current available energy for sampling during the iteration. With the advent of more accessible multi-process computing, these methods can be further expanded by combining non-uniform energy sub-domains with adaptive energy sub-domain sizing, instead of just modifying the total sampling energy domain for a single WL instance. It is worth mentioning that the systems that will benefit the most from the schemes to be outlined in the subsequent section are those that are difficult to converge. If the model system being studied rapidly converges to a desired tolerance, the benefit of parallelisation can be significantly diminished.

Of interest is also the application of GPUs for the acceleration of WL sampling, where thousands of concurrent walkers can be run on a single GPU~\cite{junqi2012parallel}. Further improvements to WL sampling on CPUs were made by the development of scalable replica exchange implementations~\cite{vogel2014replica}, with our implementation of replica exchange being discussed in the subsequent section. Another source of potential acceleration stems from the fact that $\log(g(E))$ typically follows a parabola centred on its maximum for alloys that exhibit one phase transition~\cite{pei2019fluctuation}. However, for an arbitrary system the form of $\log(g(E))$ is not known. If one knows that the underlying $\log(g(E))$ is parabolic then in theory this quantity could be initialised to a parabola or one could try to dynamically fit a parabola based upon the DOS obtained during pre-sampling. While this has potential for WL sampling, it would not apply to alloys that exhibit more than one phase transition and as such is not considered in this work.

\subsection{Parallelisation Strategies}
\label{sec:parallel_implementation}

\begin{figure*}
\centering
\includegraphics[width=\linewidth]{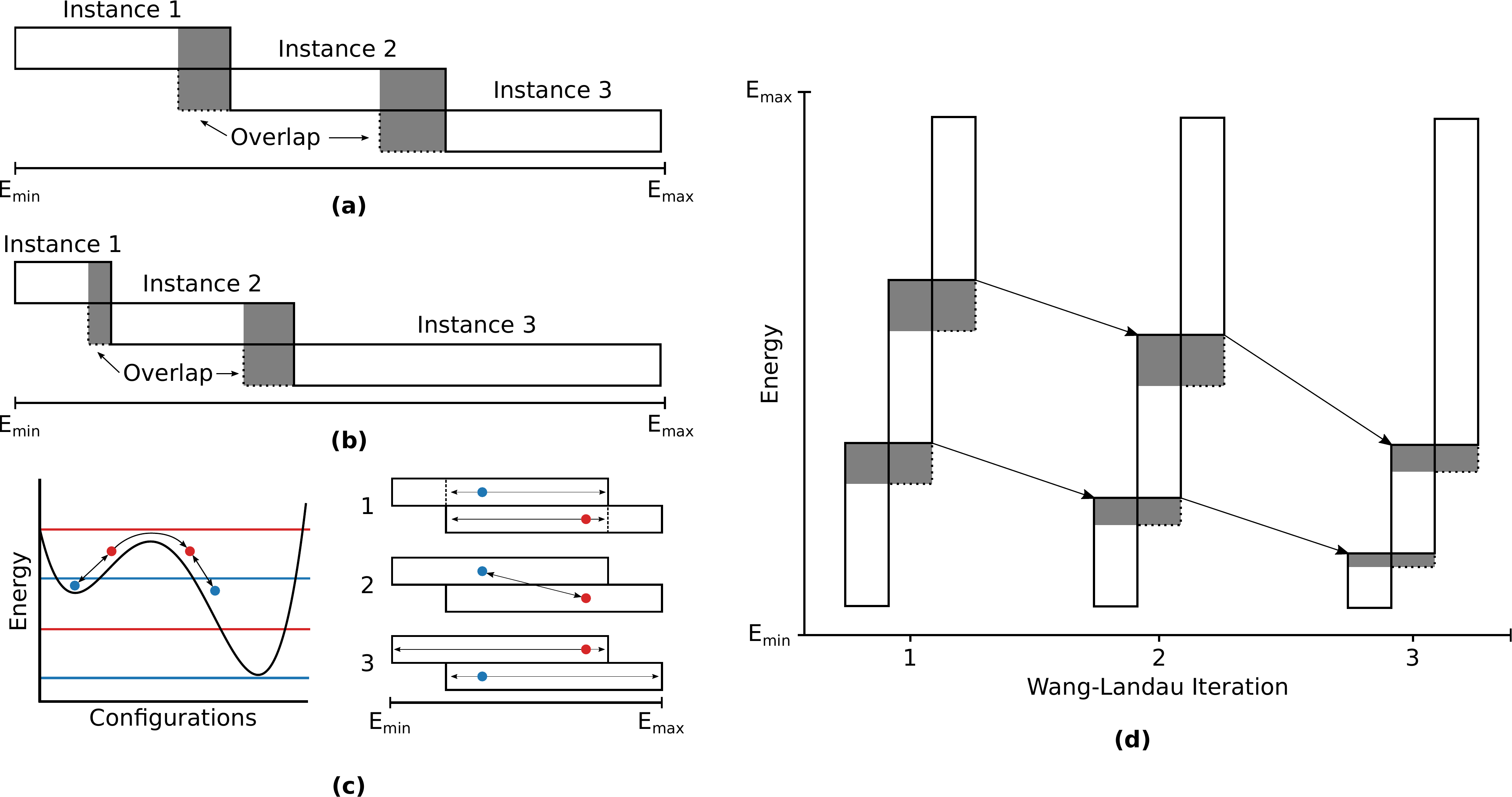}
\caption{Conceptual illustrations of the schemes discussed in this work for parallelising and/or accelerating parallel Wang--Landau sampling implementations.
    Panel (a) illustrates \textit{uniform energy domain decomposition}, where the energy domain $[E_{\mathrm{min}}, E_{\mathrm{max}}]$ is evenly partitioned into sub-domains, with fixed percentage overlapping regions (shaded). 
    Panel (b) illustrates \textit{non-uniform energy domain decomposition}, where the energy domain $[E_{\mathrm{min}}, E_{\mathrm{max}}]$ is partitioned into non-uniform sub-domains, with fixed percentage overlapping regions (shaded).
    Panel (c) illustrates \textit{replica exchange}, where independent walkers sampling within overlap regions can (occasionally) exchange configurations with neighbouring sub-domains (illustrated by red/blue particles), which allows for crossing configuration barriers. 
    Finally, panel (d) represents \textit{dynamic load balancing}, where energy sub-domains are adaptively adjusted after each Wang--Landau iteration based on the time taken to converge each sub-domain.}
    \label{fig:schemes}
\end{figure*}

In the context of parallel simulations, as mentioned in Refs.~\cite{landau_new_2004, landau_determining_2002}, it is possible to subdivide the total simulation energy domain into smaller energy sub-domains, each sampled by an independent WL simulation, also referred to as a WL instance. Within each sub-domain, multiple independent WL simulations can operate concurrently, with each separate simulation being referred to as a `random walker' or `walker'. The combination of multiple sub-domains with potentially multiple walkers per sub-domain leads to many instances of the WL algorithm. The optimal distribution of a finite number of available instances (due to finite computational resource) to walkers and windows will be explored below. It is further possible to tune the sizes of the energy sub-domains by making them occupy unequal portions of the energy domain, which is particularly useful when there are significant differences between the total number of configurations available in different energy sub-domains. These parallelism approaches have been combined with replica--exchange MC (a method that facilitates configuration exchange between walkers in neighbouring energy sub-domains)~\cite{geyer_markov_1991, hukushima_exchange_1996, valentim_exploring_2015, zhao_performance_2014} and studied previously~\cite{wl_mpi_2013}. 

Building on these ideas, all parallelisation schemes compared in this work make use of \textit{energy domain decomposition} in some form. This involves subdividing the total energy domain considered in the simulation into smaller energy sub-domains, each of which is sampled by one or more independent WL instances, \textit{i.e.} a random walker or set of walkers. In our implementation, each random walker constitutes an independent CPU process concurrently evolved throughout the simulation. A conceptual illustration of the two different energy domain decomposition schemes considered in this work is given in panels (a) and (b) of Figure~\ref{fig:schemes}. It should also be stressed that implementation complexity is a key factor to consider when selecting parallelisation scheme(s). Therefore, below, once the parallelisation schemes have been outlined, we shall discuss this key aspect.

When the simulation begins, each energy sub-domain is assigned a walker (or set of walkers) and is initialised with a given physically relevant system configuration. (In the context of the alloys studied in this work, each walker is initialised with a random, lattice-based atomic configuration.) Frequently, these initial configurations have energies which lie outside the energy sub-domain assigned to the WL instance. Before sampling can begin, these walkers must be evolved so that they lie in the assigned energy sub-domain. In this work, this is achieved by running an MC simulation with an acceptance probability of
\begin{equation}
    P_{n\rightarrow m} = 
    \begin{cases}
        \; 1, &\quad (E_m - E_T)^2 - (E_n - E_T)^2 \leq 0 \\
        \; 0, &\quad (E_m - E_T)^2 - (E_n - E_T)^2 > 0,
    \end{cases}
\label{eq:burn_probability}
\end{equation}
where $E_T$ is the `central' energy of the target sub-domain. In complex energy landscapes, it remains possible for a walker to become `trapped' in a region of configuration space. To remedy this potential issue in the present study, if a walker is not within its specified energy sub-domain following a chosen number of trial MC moves, its atomic configuration is reinitialised to a random state and re-evolved towards its targeted energy using the acceptance probabilities defined in Eq.~\ref{eq:burn_probability} once more. We note that this approach functions in the case of the present work, but might not necessarily be broadly applicable. In other contexts with more complex energy landscapes it may be appropriate to apply a harmonic umbrella potential~\cite{kastner2011umbrella} to each walker to encourage it toward its assigned window during this initial ``burn-in'' phase of the simulation. 

Each energy sub-domain overlaps with its adjacent energy sub-domains to facilitate combining the local DOS (the DOS associated with each energy sub-domain) into a global DOS across the whole domain. After each WL iteration has concluded, the global DOS is pieced together starting with the lowest energy sub-domain. The first and second energy sub-domains have their DOS combined by finding where the statistical temperature
\begin{equation*}
    \beta = \frac{\textrm{d}\log\left[g\left(E\right)\right]}{\textrm{d}E}
    \;\approx\; \frac{\log g(E+\Delta E) - \log g(E)}{\Delta E},
\end{equation*}
computed from each local DOS, best coincides within the overlap region. This is done by calculating $\beta$ for each pair of neighbouring bins within the overlap region and then selecting the $\beta$ that best coincides between the two neighbouring energy sub-domains as the joining point for the DOS. This process is repeated for each subsequent energy sub-domain until a pieced together DOS is obtained for the whole energy domain. If there are multiple walkers within an energy sub-domain, their individual DOS are averaged before the DOS are combined.

Due to the division of the energy domain into multiple sub-domains, and non-uniformity of simulation convergence times across these various sub-domains, additional modifications are made to determine when a WL iteration is finished. The modification factor remains constant across all sub-domains during an iteration and is only updated once the WL iteration ends. All walkers continue sampling with the current modification factor, even if they have satisfied the flatness criterion, until all walkers within the simulation also meet the flatness criterion. This avoids wasted CPU time by preventing WL instances from idling while `waiting' for other walkers to converge.

\subsubsection{Choice of Energy Domain Overlap Size}

In the case of uniform energy domain decomposition, construction of energy sub-domains begins with uniformly dividing the energy domain into equal-sized sub-domains, after which the overlap is determined as a fixed percentage of the preceding sub-domain. This creates an overlap between an energy sub-domain and both of its neighbours.

The overlap calculation proceeds from lower to higher energy sub-domains. The first sub-domain is kept at its original size. Each subsequent sub-domain (sub-domain $i+1$) is extended to include a fixed percentage of the updated size of the preceding sub-domain (sub-domain $i$), rounded up to the nearest integer number of energy bins. In the present study, we always require a minimum overlap of 2 bins between neighbouring energy sub-domains to facilitate reliable construction of the global DOS.

\subsubsection{Non-Uniform Energy Domain Decomposition}
\label{subs:non-uniform}
The time taken for each WL iteration is given by the time to converge of the slowest-converging sub-domain, since each walker must first converge their local (sub-domain) DOS histogram before the WL iteration can finish and create a combined global DOS. In general, the time taken to converge each sub-domain is non-uniform across the total energy domain, tending to take longer to converge around phase transitions or at low energy in the presence of degenerate microstates. In the former case, systems may exhibit entropic bottlenecks which are difficult to sample, and in the latter case the number of accessible configurations necessitates long sampling times. This non-uniform convergence time can be somewhat mitigated by altering the size of each WL energy sub-domain, however optimal load balancing is impossible \textit{a priori}, as that requires the converged DOS states, which is the quantity being computed via simulation. By adjusting the individual size of each sub-domain, one can obtain an energy decomposition which is non-uniform in energy. This is done before the first iteration via a pre-sampling step, which evolves the WL algorithm until each energy bin across the entire energy domain has been visited by at least one walker per sub-domain a thousand times. We selected this number as a compromise between excessive simulation time being spent in the pre-sampling step and ensuring that every bin has sufficient visits such that the initial DOS estimate is reasonable. We find that the selection of this parameter is not crucial, since an estimate of the DOS must be built up from an initial state of $\log(g(E))$ which takes an appreciable portion of simulation time regardless, and so the parameter is chosen to err on the side of caution to ensure that all bins are appropriately sampled. Performing a pre-sampling step that is much longer than this is inefficient, since the simulation is progressing with sub-domain sizes which are non-optimal. This pre-sampling step is performed with uniformly sized sub-domains and lasts on the order of one WL iteration. The time taken per sub-domain is used to obtain the initial non-uniform energy domain decomposition. The details of the explicit means by which this initial non-uniform energy domain decomposition is obtained are discussed below, in Sec.~\ref{sec:load_balance}. A conceptual illustration of a non-uniform energy domain decomposition scheme is given in Figure~\ref{fig:schemes}, panel (b).

\subsubsection{Replica Exchange}

Due to the often complex nature of the configuration space and associated energy landscape of the alloy being studied, replica exchange has been implemented within our test case to investigate its effectiveness in the context of the present work. Replica exchange allows for energy sub-domains to exchange walkers with neighbouring energy sub-domains provided that both configurations have energies which lie within the shared overlap region. This can, in principle, provide a way for walkers to reach potentially inaccessible regions within their assigned energy sub-domain by crossing energy barriers that require higher energies to access. This is anticipated to be particularly useful at low energies, where it could prevent walkers lying within a low-energy sub-domain from being trapped within a local minimum of the energy landscape. A conceptual illustration of replica exchange is given in Figure~\ref{fig:schemes}, panel (c). Within the model simulation, replica exchange swaps are proposed every 10 MC sweeps---where each MC sweep consists of a number of MC moves equal to the total number of atoms in the simulation cell---and the replica exchange swap is accepted based on acceptance probability of Eq.~\ref{eq:wl_transition_probability}, \textit{i.e.} the same acceptance probability as for any regular trial MC move.

\subsubsection{Dynamic Load Balancing}
\label{sec:load_balance}

As mentioned above, in Sec.~\ref{subs:non-uniform}, perfectly optimal load balancing is anticipated to be impossible in general for arbitrary model systems. However, one can begin to approach optimal load balancing by iteratively adjusting the size of energy sub-domains after each complete WL iteration based on the time taken to converge the individual sub-domain visit histograms. A conceptual illustration of how such an iterative scheme might function is given in Figure~\ref{fig:schemes}, panel (d).

In this work, we propose a dynamic load balancing scheme which adjusts the sizing of energy sub-domains based on the run-time(s) of the walker(s) operating in that sub-domain at the end of the current WL iteration. This dynamic adjustment occurs after each WL iteration, and is carried out using the same algorithm as was proposed in the pre-sampling step for non-uniform energy domain decomposition, Sec.~\ref{subs:non-uniform}. The number of energy bins, $d_i$, assumed to be the optimal number to be included in energy sub-domain, $i$, for a given iteration, $k$, is computed simply based on the relative MC steps taken to converge each sub-domain in that iteration according to 
\begin{equation}
    d_{i,k} = \frac{1/t_{i,k}}{\sum_i 1/t_{i,k}} N,
\end{equation}
where $i$ is the index of the energy sub-domain, $t_{i,k}$ is the number of MC steps taken per bin for energy sub-domain $i$ to reach the flatness criterion in iteration $k$, and $N$ is the total number of energy bins across the entire energy domain.

Our proposed extension of this scheme also updates the sub-domain size at each subsequent WL iteration according to
\begin{equation}
\begin{aligned}
    &d_{i, k+1} = a_k\cdot d_{i,k-1} + b_k\cdot d_{i,k} \\[1em]
    &a_k + b_k = 1
\end{aligned}
\end{equation}
where $d_{i,k}$ is the size of sub-domain $i$ after iteration $k$ as a fraction of the total number of bins, and $a_k$ and $b_k$ are scaling factors. To ensure the entire energy domain is covered, $\sum_i d_{i,k+1} = N$. The purpose of $a_k$ and $b_k$ is to introduce memory (\textit{i.e.} incremental changes to previous sub-domain sizing) into the size of the energy sub-domains in order to reduce overcompensating after an outlier in convergence time. With each subsequent WL iteration $k$, $b_k$ decreases such that the sub-domain adjustment decreases with each iteration. Explicitly, in this work, $a_k$ and $b_k$ were chosen according to
\begin{equation}
\begin{aligned}
    a_k &= 1 - b_k, \\
    b_k &= \alpha \cdot \beta^k, \\
    a_0 &= 1 - b_0, \\
    b_0 &= \alpha \cdot \beta^0,
\end{aligned}
\end{equation}
where $\alpha$ controls the memory, and $\beta$ controls the adjustment rate of sub-domain size. We find $\alpha=0.2$ and $\beta=0.9$ to be a reliable choice of parameters for use in the context of the present work. While results are sensitive to the choice of these parameters, their only effect is the rate at which the optimal domain sizes are found. In general, one should take care to choose $\alpha$ and $\beta$ such that the rate of change of domain size decreases appreciably to prevent excessive oscillation of domain size. Of note, the initial update following the pre-sampling step is performed using a uniformly initialised $d$ with $b=1$. Additionally, in this work, to prevent zero-bin energy sub-domains, a minimum sub-domain size is imposed of 2\% of the total number of bins across the entire energy domain.

After such a sub-domain adjustment, walkers may no longer be in their designated sub-domain and as such they are moved into their appropriate energy domain, as outlined at the start of this section, \textit{i.e.} by performing (and restarting as necessary) an MC simulation with acceptance probability as defined in Eq.~\ref{eq:burn_probability}. (Note that accumulation of the histogram(s) and refinement of the DOS are paused during this adjustment process.) The sizes of energy overlap regions are also re-computed for the adjusted sub-domain sizing. We note here that a related approach has previously been explored, in which a single load balancing step was performed following the pre-sampling step~\cite{wl_mpi_2013}, but where there was no further adjustment of energy sub-domain sizing following this initial load-balancing. We stress however, that this per-iteration adjustment has negligible computational overhead as compared to the total runtime of the simulation ($<0.1\%$, taking $\sim$15~ms per call in our implementation).

\begin{figure*}[!t]
    \centering
    \includegraphics[width=\linewidth]{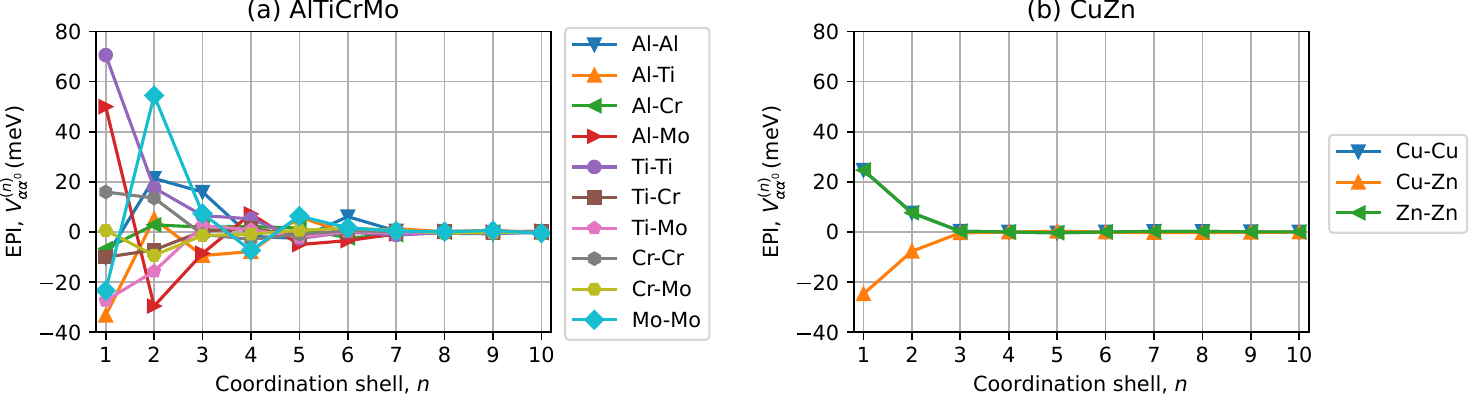}
    \caption{Plots of real-space effective pair interactions, $V_{\alpha \alpha'}^{(n)}$, as a function of coordination shell number, $n$, recovered from the reciprocal space $S^{(2)}_{\alpha \alpha'}({\bf k})$ data for (a) AlTiCrMo and (b) CuZn. The notation $V_{\alpha \alpha'}^{(n)}$ indicates the interaction between chemical species $\alpha$ and $\alpha'$ on coordination shell $n$, \textit{i.e.} at $n$\textsuperscript{th} nearest neighbour distance. For both alloys, it can be seen that the comparative strength of interactions tails off quickly with increasing distance and is particularly rapid for CuZn. Note that, for the CuZn binary, the Cu-Cu and Zn-Zn effective pair interactions are equal on all coordination shells due to the choice of gauge~\cite{khan_statistical_2016}.}
    \label{fig:epi_comparison}
\end{figure*}

\subsubsection{Implementation Complexity}

We now comment on the implementation complexity of each of the outlined parallelisation schemes and features. We highlight that, provided all parallelism features are implemented one after another, they form natural extensions of each other, simplifying the overall implementation. In order, the complexity of implementation of the various features are summarised as follows:
\begin{itemize}[leftmargin=8pt]
    \item \textbf{Energy domain decomposition (uniform).} Implementation of uniform domain decomposition is straightforward as each sub-domain can be treated as an independent WL instance with its own histograms, and therefore requiring no changes to the core WL algorithm. The two main implementation requirements are: ensuring all WL instances share the same modification factor and flatness criterion, and enforcing that each WL instance remains within its assigned energy interval.
    \item \textbf{Global DOS construction.} Constructing a global DOS is the primary non-trivial component of uniform energy domain decomposition. A robust combination procedure is needed that is capable of identifying a joining point within the overlap region between neighbouring sub-domains (here based on $\beta(E)$), and applying a consistent offset between $\log g(E)$ segments so that they match at the selected joining point. If this is performed without sufficient care, artefacts in the resulting DOS can occur at the boundaries between sub-domains.
    \item \textbf{Energy domain decomposition (non-uniform).} Non-uniform energy domain decomposition is more complex than the uniform case but retains the same overall structure. Each sub-domain remains as an independent WL instance and, as such, most uniform decomposition functionality can be reused with the only additional requirement being a method for measuring convergence time per sub-domain. These timings can then be used during a pre-sampling step to define new, non-uniform sub-domains, after which the existing functionality for keeping instances within their target domains can be reused.
    \item \textbf{Replica exchange.} Replica exchange has moderate implementation complexity. It introduces explicit communication between neighbouring sub-domains via configuration swaps, requiring a synchronised swap protocol (proposal schedule, overlap energy eligibility checks, and accept/reject decisions). Care is needed to avoid communication deadlocks and to ensure exchanges occur only between eligible WL instances. Unlike overlap handling or DOS construction, replica exchange must be integrated directly into the sampling loop rather than added as an intermediate step between WL iterations.
    \item \textbf{Dynamic load balancing.} While dynamic load balancing appears complex, it becomes a straightforward extension of non-uniform domain decomposition once this feature is implemented. The additional functionality required is a per-iteration mechanism to update sub-domain boundaries, and a method to map each updated sub-domain onto the corresponding region of the global DOS.
\end{itemize}
For the purposes of the present work, we have implemented these schemes in the open-source \textsc{BraWl} package~\cite{naguszewski_BraWl_2025}. 

\subsection{Model Implementation}
\label{sec:bw_model_implementation}

As a test case for the parallelisation strategies studied in this work, we consider the application of WL sampling to the study of the phase behaviour of two selected metallic alloys: one well-studied binary system~\cite{khan_density-functional_2016} and one multicomponent, high-entropy alloy (HEA). The application of WL to multicomponent alloys is an emerging area of research~\cite{pei_statistics_2020, takeuchi_new_2017, nanba2023implementation} and is therefore a relevant test case for this study, while the binary system provides a simple test case to compare against the HEA.

HEAs are an intriguing class of disordered material, typically containing a mixture of four or more metallic elements combined in near-equal ratios, and are of interest for a range of next-generation technological applications as they have been shown to frequently possess superior physical properties when compared to traditional binary/ternary alloys. The large contribution made to the free energy of these materials by the configurational entropy is generally understood to stabilise disordered `solid solutions'~\cite{yeh_nanostructured_2004}, in which there is a fixed, underlying crystal lattice, but where the constituent elements of the material occupy these lattice sites in a random fashion. Several of such systems have been shown to exhibit both partial crystallographic ordering and/or phase decomposition with decreasing temperature. Computational modelling can shed insight into such diffusional solid-solid phase transformations. However, in the context of HEAs, computational simulation of phase equilibria faces several key challenges~\cite{widom_modeling_2018, eisenbach_first-principles_2019, ferrari_frontiers_2020, ferrari_simulating_2023}. Primary among these is that the space of possible configurations (henceforth referred to as the `configuration space') grows combinatorially both with the number of elements present in a considered composition, as well as with the number of atoms in a given simulation cell~\cite{zhang_roadmap_2025}. This means that a large number of configurations must be sampled for there to be confidence that results are well-converged and recovered quantities are representative of the relevant thermodynamic phases. WL sampling is hence extremely valuable in this context. 

In this work, we consider the application of WL sampling to a fixed lattice model describing the internal energy of a substitutional alloy. As benchmark systems, we select the AlTiCrMo high-entropy alloy, as well as the binary CuZn system. AlTiCrMo is classified as a refractory high-entropy superalloy~\cite{miracle_refractory_2020} and forms a (partially) chemically ordered structure with decreasing temperature~\cite{chen_crystallographic_2019}. In the context of the present work, AlTiCrMo represents a complex system with competing interactions, multiple phase transitions, and a wide range of possible ordered configurations~\cite{woodgate_emergent_2025}. By contrast, CuZn is a well-studied binary alloy which unambiguously undergoes a single phase transition from a disordered bcc (A2) structure to an ordered B2 structure at intermediate temperatures~\cite{khan_density-functional_2016}. It thus represents a simpler system than AlTiCrMo and allows us to examine how our proposed parallelisation strategies perform on model systems of varying complexity.

The parallelisation schemes considered in this study are implemented within the \textsc{BraWl} package~\cite{naguszewski_BraWl_2025}, which focuses specifically on fixed-lattice simulations of substitutional alloys. In particular, the code implements the Bragg--Williams model~\cite{bragg_effect_1934, bragg_effect_1935}, which describes the internal energy of a given alloy as a sum over atom-atom effective pair interactions, which are assumed to have finite range. Explicitly, this Hamiltonian takes the form 
\begin{equation}
    H(\{\xi_{i\alpha}\}) = \frac{1}{2}\sum_{i \alpha; j\alpha'} V_{i\alpha; j\alpha'} \xi_{i \alpha} \xi_{j \alpha'},
    \label{eq:b-w1}
\end{equation}
where $V_{i\alpha; j\alpha'}$ denotes the effective pair interaction (EPI) between an atom of chemical species $\alpha$ on lattice site $i$ and an atom of chemical species $\alpha'$ on lattice site $j$. $\{\xi_{i\alpha}\}$ indicates the site occupancies, where $\xi_{i\alpha}=1$ if site $i$ is occupied by an atom of chemical species $\alpha$, and $\xi_{i\alpha}=0$ otherwise. It is helpful to observe that this Hamiltonian can be thought of as a generalisation of the Lenz--Ising model used in elementary studies of magnetic phase transitions~\cite{brush_history_1967}. The EPIs of Eq.~\ref{eq:b-w1} are generally assumed to be homogeneous, isotropic, and of finite range, which simplifies evaluation of this expression. In this work, the EPIs used are obtained using the $S^{(2)}$ theory for multicomponent alloys, which is discussed extensively in earlier works~\cite{khan_statistical_2016, woodgate_compositional_2022, woodgate_modelling_2024}. For AlTiCrMo, we take the EPIs previously calculated in Ref.~\cite{woodgate_emergent_2025}, which include interactions between pairs of atoms up to sixth-nearest neighbour distance and can be seen in panel (a) of Figure~\ref{fig:epi_comparison}. For CuZn, we calculate EPIs at an optimised lattice constant of $a=2.971$~\AA~for the disordered bcc (A2) phase obtained using density functional theory (DFT) calculations with the generalised gradient approximation (GGA) to the exchange-correlation functional as parametrised by Perdew, Burke, and Ernzerhof~\cite{perdew_generalized_1996}. The obtained EPIs can be seen in panel (b) of Figure~\ref{fig:epi_comparison}. The Korringa--Kohn--Rostoker method is used~\cite{ebert_calculating_2011}, with the coherent potential approximation~\cite{soven_coherent-potential_1967} employed to average over chemical disorder. In contrast to AlTiCrMo, for CuZn the EPIs are found to be much shorter in range and only significant on the first two coordination shells of the bcc lattice. We therefore truncate the EPIs for CuZn at this range. 

\subsection{Performance Metrics}
\label{sec:performance_metrics}

\begin{table*}[!ht]
    \centering
\begin{ruledtabular}
\begin{tabular}{cccc}
Method Label & \multicolumn{3}{c}{Parallelism Scheme(s)}                       \\ \hline
1            & Dynamic Load Balancing & Non-Uniform Sub-domains & Replica Exchange \\
2            & Dynamic Load Balancing & Non-Uniform Sub-domains &                  \\
3            &                        & Non-Uniform Sub-domains & Replica Exchange \\
4            &                        & Non-Uniform Sub-domains &                  \\
5            &                        & Uniform Sub-domains & Replica Exchange \\
6            &                        & Uniform Sub-domains     &                 
\end{tabular}
\end{ruledtabular}
    \caption{Parallelism methods and associated labels used within the present study. All available combinations of the three separate parallelism methods are listed. Of note, it is not possible to have dynamic load balancing unless the energy sub-domains are non-uniform, as dynamic load balancing has non-uniform energy sub-domain sizing as a pre-requisite.}
    \label{tab:methods}
\end{table*}

In order to quantify the relative performance of the parallelisation schemes studied in this work, the following performance metrics are defined. First, we define the speedup, $S$, via
\begin{equation}
    S\left(h,m\right) = \frac{N(1,m)}{N(h,m)},
\end{equation}
where $N(h,m)$ is the time taken to reach the flatness criterion by the slowest energy sub-domain, $h$ is the total number of energy sub-domains, and $m$ is the number of walkers per window. 

We then define two metrics for measuring simulation efficiency. The first of these, $\phi_h$, is used to quantify sub-domain efficiency and is defined as
\begin{equation}
    \phi_h\left(h,m\right) = \frac{N(1,m)}{h \cdot N(h,m)},
\end{equation}
while the second, $\phi_m$, is used to quantify walker efficiency and is defined as
\begin{equation}
    \phi_m\left(h,m\right) = \frac{N(1,m)}{m \cdot N(h,m)}.
    \label{eq:walker_efficiency}
\end{equation}
The sub-domain efficiency, $\phi_h$, measures the mean speedup per sub-domain with a fixed number of walkers per domain, whereas the walker efficiency, $\phi_m$, measures the mean speedup per walker at fixed number of sub-domains. In previous work~\cite{wl_mpi_2013}, efficiencies of greater than 100\% have been observed, and we expect the same to be observed here. Though initially counter-intuitive, efficiencies greater than 100\% are possible in this context since not only does domain decomposition divide the work over concurrent CPU processes, it also has the potential to reduce the total number of energy evaluations required to converge the WL sampling simulation. This reduction in the work needed can be distinguished from traditional measures of efficiency in the context of CPU parallelism, since this reduction in work and hence speedup from sub-domain decomposition can be observed even on a single CPU sampling multiple sub-domains in order~\cite{cunha-netto_improving_2008}. 

We can demonstrate approximately how the time taken for a sub-domain to converge is anticipated to scale with energy sub-domain size using a simple argument based on dimensional analysis. For a single walker, we model the energy sampled over time as a diffusive random walk in energy space. The time it takes for the probability distribution of finding the walker at energy $E$ to evolve from an initial Dirac delta function to an eventual uniform distribution can be inferred from dimensional analysis as
\begin{equation*}
    t \propto D^\alpha\cdot L^\beta,
\end{equation*}
where $t$ is the `time' (\textit{i.e.} number of MC trial moves) taken for a walker to diffuse across the sub-domain, $D$ is the diffusivity and $L$ is the size of the sub-domain. Using dimensional analysis one can obtain $\alpha = -1$ and $\beta = 2$ which leads to
\begin{equation*}
    t \propto \frac{L^2}{D}
\end{equation*}
from which we can see that $t \propto L^2$ for any fixed diffusivity, $D$. As such, the time taken for a walker to explore a sub-domain scales quadratically with the size of that sub-domain. In the case of $h$ uniformly sized sub-domains the time taken, $t_h$, to converge is thus proportional to $\left(\frac{L}{h}\right)^2$ per domain. Assuming that the same constant of proportionality holds for both the total energy case and the case of $h$ uniform sub-domains, we have that $t_h/t = 1/h^2$, and it follows that 
\begin{equation}
    ht_h = \frac{t}{h},
\end{equation}
\textit{i.e.} the total simulation time when the total energy domain is subdivided into $h$ energy domains, $ht_h$, is reduced by a factor of $h$ compared to the simulation time when running across the total sub-domain alone. Within this simple model, simulating each sub-domain on a separate CPU process can in principle result in a $h^2$ speedup and a sub-domain efficiency of $h \cdot 100\%$. However, this is a highly idealised model which does not account for overlaps between sub-domains, variation in diffusivity $D$ across the energy domain and communication overheads. There is also an upper limit to $h$, beyond which there is an insufficient number of bins per sub-domain for the algorithm to function, motivating the use of multiple walkers. In principle however, the use of non-uniform sub-domain sizes should exactly compensate for the non-uniformity of $D$ and hence it should be possible to approach the ideal speedup of $h^2$ provided the size of overlap regions, as well as practical details such as parallel communication overheads, are both minimised.

\section{Results and Discussion}
\label{sec:results}

In order to investigate the various outlined parallelisation schemes, the possible parallelisation schemes are grouped and labelled as `methods', as shown in Table~\ref{tab:methods}. Once grouped, the effect of the schemes on speedup per WL instance, efficiency per sub-domain and efficiency per walker, was investigated over a range of $h$ and $m$. The specific model systems being simulated were the equiatomic AlTiCrMo HEA, the phase behaviour of which has been investigated in an earlier work ~\cite{woodgate_emergent_2025}, and the equiatomic CuZn alloy. The phase behaviour of these two systems is described in the Supplementary Material. We shall first begin by addressing the more complex of the two alloys, namely AlTiCrMo.

\subsection{{AlTiCrMo}}

In this benchmarking study assessing the efficiency of various parallelisation schemes, we use a simulation supercell made up of $4 \times 4 \times 4$ bcc cubic unit cells for a total of 128 atoms. We find that 128 atoms is sufficient to represent the sampling challenge with tests at larger lattice sizes ($6\times6\times6$ bcc cubic unit cells, 432 atoms) confirming the trends illustrated here. Our WL sampling simulations used $N=512$ energy bins over an energy domain of $E=-99$~meV/atom to $E=0$~meV/atom (obtained via the method outlined in Sec.~\ref{sec:wl_sampling} with $T_{\text{min}}=50K$ and $T_{\text{max}}=3000K$), completing 12 full WL iterations starting with $\log f_1=0.05$, where each iteration used half of the logarithm of the modification factor of the previous iteration. The AlTiCrMo HEA model is well-suited as a benchmark system as it exhibits two phase transitions which occur at two separate temperatures~\cite{woodgate_emergent_2025}, and hence convergence of the DOS is non-trivial. We investigated performance for a range of parameters within our parallelisation schemes with the number of walkers per sub-domain $m$ varying from 1 to 6, the number of sub-domains $h$ varying from 1 to 16 and the sub-domain overlap varying from 0\% to 75\% for both overlap schemes. It is important to note that we impose a requirement that a minimum of two bins must lie in the overlap region. As such, the case which we refer to as 0\% overlap still has two bins in the overlap region. This minimum overlap does not change based on the sub-domain size. 

Throughout, we plot results as a function of the total number of WL instances ($m\times h$) used in the calculation. We emphasise once again here that, in this work, evaluation of the model Hamiltonian (Eq.~\ref{eq:b-w1}) is performed on a single CPU core, and hence the number of CPU cores strictly refers to the number of parallel WL simulation instances. If a more complex model Hamiltonian were to be studied, requiring multiple cores to be utilised for evaluation of the Hamiltonian, the total core count for a simulation would be $m\times h \times n_H$, where $n_H$ was the number of cores used for evaluation of the model Hamiltonian. We therefore assert that our results will remain relevant for larger-scale simulations than those presented here.

Where error bars are presented on quantities, these are obtained by running five independent WL sampling simulations using different seeds on the pseudorandom number generator (PRNG) for the given choice of parallelisation scheme and associated settings. The data points represent the obtained average, while the error bars represent the standard deviation calculated across the five samples. We note that the default PRNG used in \textsc{Brawl}~\cite{naguszewski_BraWl_2025} uses the Mersenne Twister algorithm~\cite{matsumoto_mersenne_1998}, which has a long period and is well-suited to long sampling runs.

\subsubsection{Effect of energy sub-domain overlap}

\begin{figure}[t]
    \centering
    \includegraphics[width=\linewidth]{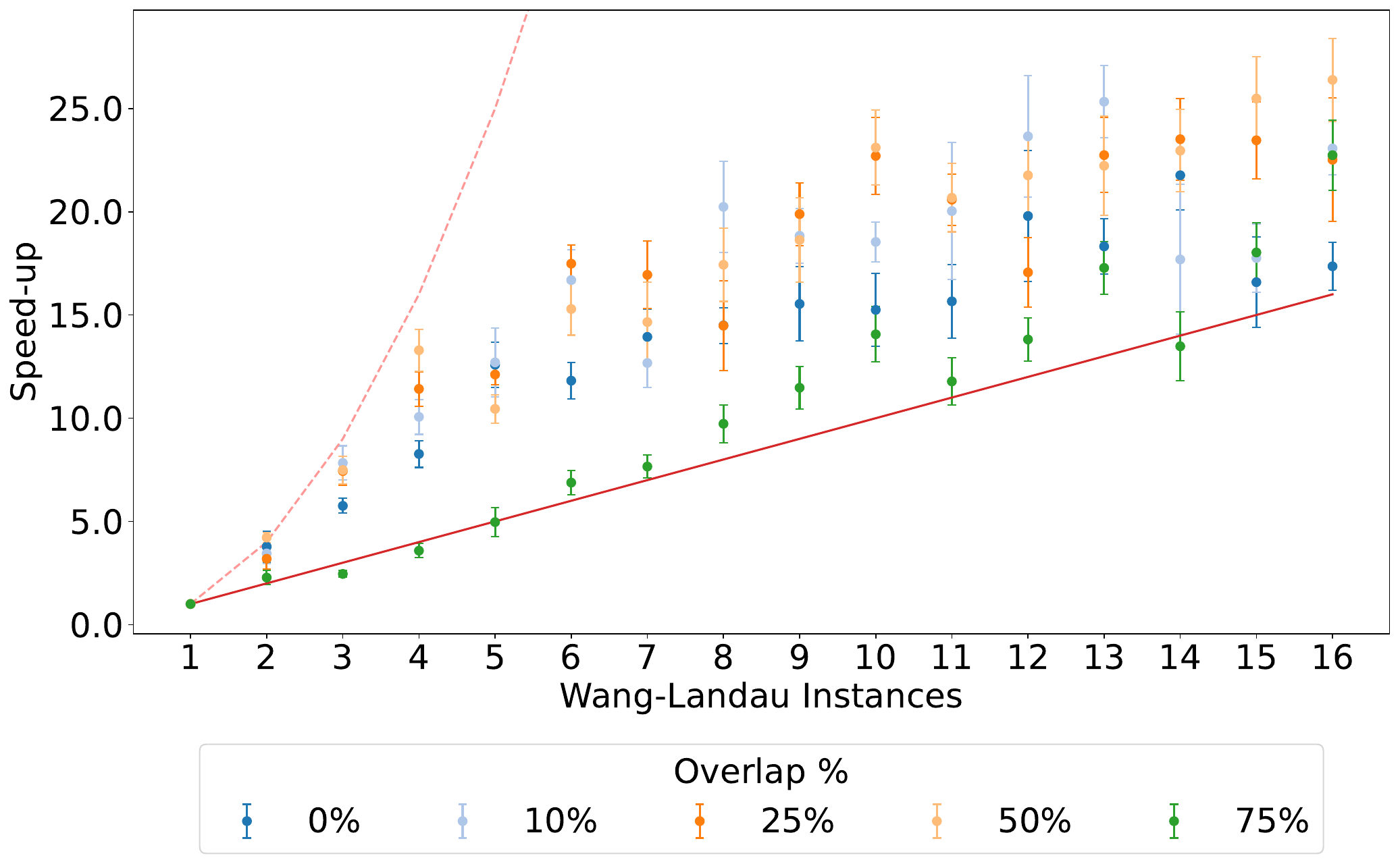}
    \caption{Comparison of the speedup per WL instance for different choices of energy sub-domain overlap, taken relative to the case of a single walker across the entire energy domain for simulations of the AlTiCrMo test system. This plot is for simulations using Method 3 with a single walker per energy sub-domain. The dotted red line shows the square of the number of energy sub-domains, \textit{i.e.}, the number of WL instances used squared ($h^2$), while the solid red line shows the number of energy sub-domains, \textit{i.e.}, the number of WL instances ($h$). It can be seen that, for all choices of energy sub-domain overlap region size (excluding the 75\% case), the speedup is significantly above 100\% and that the choice of size of overlap region between energy sub-domains has noticeable impact on speedup.}
    \label{fig:overlap_nrmse_speedup}
\end{figure}

We begin by considering the effect of energy sub-domain overlap. We choose to use Method 3 to illustrate this effect. We also choose to employ only a single walker per energy sub-domain, as we do not expect the number of walkers per sub-domain to change the impact of overlap region size on simulation speedup. Figure~\ref{fig:overlap_nrmse_speedup} compares the obtained simulation speedup as a function of number of WL instances employed for several choices of energy sub-domain overlap region size. It can be seen that the degree of overlap does not have a significant impact on the speedup achieved through WL sampling. Across all sub-domain counts, only in the 75\% overlap case does the speedup begin to reduce significantly relative to the other choices of overlap region size.

Of the overlaps, the 25\% to 50\% overlap cases appear to be the ones with the most speedup, although not by a statistically significant amount. Such sub-domain overlap allows for better combining of the DOS, as well as a larger region for replica exchange, while still allowing walkers to focus more narrowly on their assigned sub-domain, resulting in more efficient sampling and higher speedup. The other cases introduce either too much or too little overlap, with too little overlap potentially harming the DOS combining and too much overlap leading to a greater portion of the configuration space being sampled by multiple walkers, which reduces the benefits of parallelisation. For the remainder of this study a 25\% overlap is used. This deliberate choice allows for evaluation of performance with an overlap width which remains optimal to within the statistical uncertainty of our data, but still provides a significant range of overlap to realise any benefit from including replica exchange.

It is important to note the super-linear scaling observed in some cases. This greater than 100\% speedup per WL instance arises from the reduced number of total MC steps required when sampling due to the energy domain decomposition as explained in the previous section. This effect was discussed above, in Sec.~\ref{sec:performance_metrics}, and is also examined further in Sec.~\ref{sec:super-linear}, below. Of further note is the speedup for the first 3 WL instances matches the theoretical limit of $h^2$, derived in Sec.~\ref{sec:performance_metrics} to within the uncertainty of our measurements.

\begin{figure}[t]
    \centering
    \includegraphics[width=\linewidth]{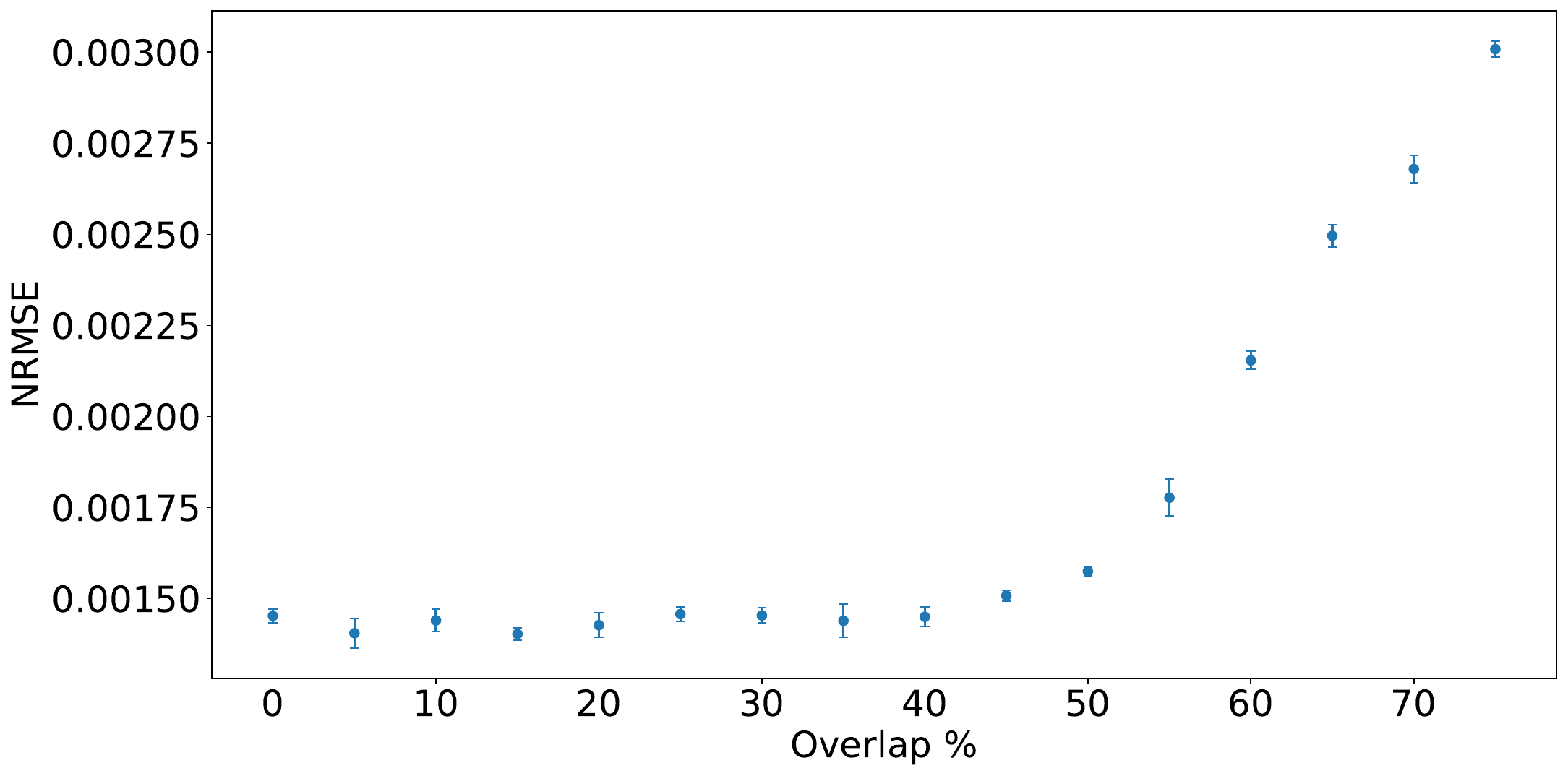}
    \caption{Normalised root mean square error (RMSE) as a function of the percentage overlap for simulations of the AlTiCrMo test system. The RMSE was obtained by comparing 23 iteration sampling runs with 16 windows to a well-converged 36 iteration sampling run with 1 window. The error bars correspond to the standard deviation of time taken of 5 repeat WL simulations. For each number of overlap bins, 5 separate 23 iteration simulations were performed and averaged. These data show that there is no correlation between the size of the energy sub-domain overlap region and the accuracy of the obtained global DOS up to a 50\% overlap. Beyond 50\% overlap the averaging power of the sampling is reduced hence the increase in normalised RMSE though not to a statistically significant extent.}
    \label{fig:overlap_nrmse}
\end{figure}

To assess the effect of overlap on DOS convergence, in Figure~\ref{fig:overlap_nrmse} we show the normalised root mean square error (RMSE) between a 36 iteration (well-converged) WL simulation in one sub-domain and 23 iteration WL simulations of varying overlap size with 16 sub-domains. It can be seen that there is no correlation between DOS convergence accuracy and the percentage overlap in the energy sub-domain overlap region up to 50\%. Beyond this point the averaging power of the overlap regions is decreased resulting in an increased normalised RMSE. Of note, however, is that this increase in normalised RMSE increases from 0.15\% to 0.3\% which is not a statistically significant increase. While remaining at or below a 50\% overlap, the primary benefit of the energy sub-domain overlap region(s) is to facilitate replica exchange and does not harm the sampling.

\subsubsection{Influence of number of walkers used per energy sub-domain}

\begin{figure}[t]
    \centering
    \includegraphics[width=\linewidth]{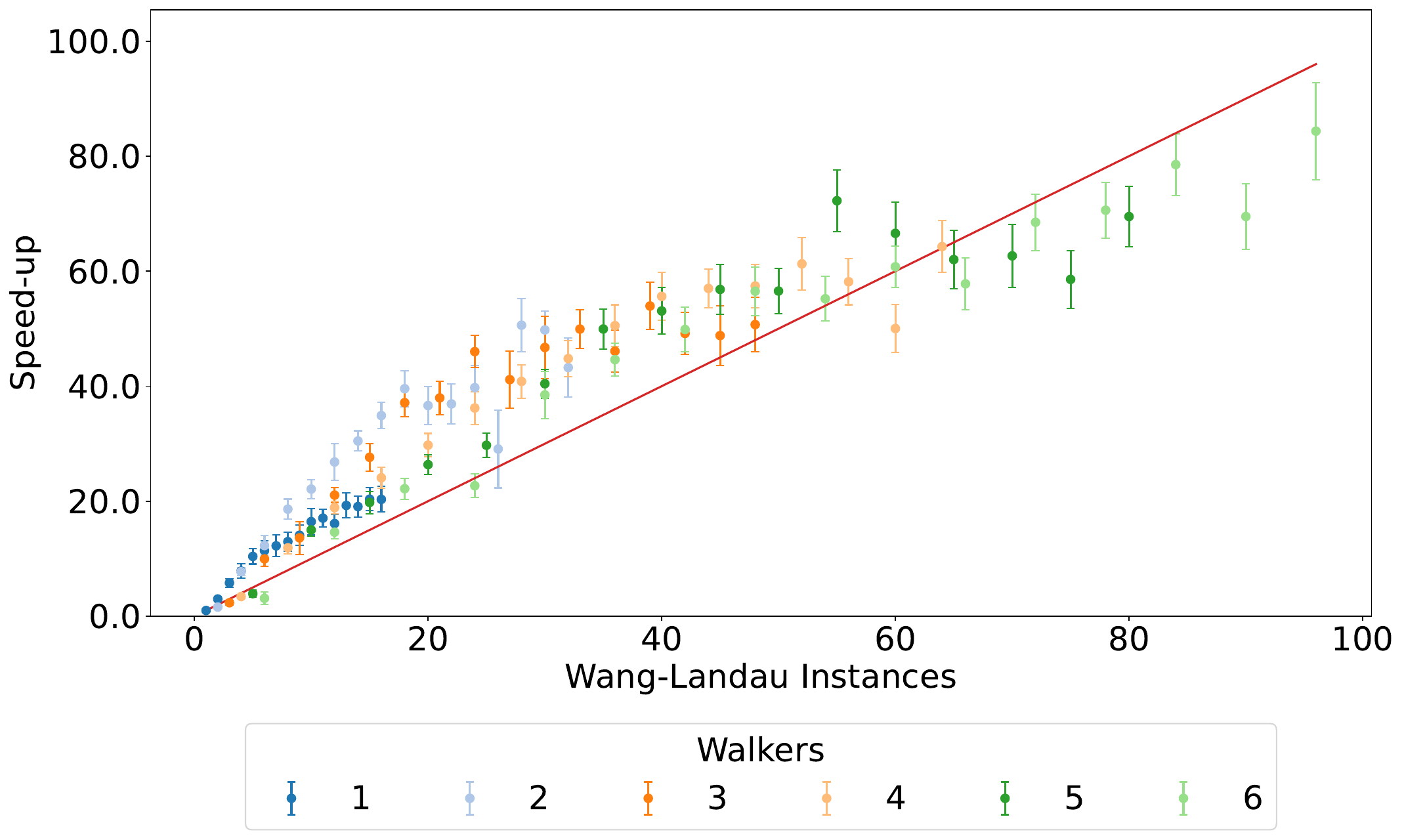}
    \caption{Comparison of the speedup per WL instance for different numbers of walkers per energy sub-domain, taken relative to the case of a single walker across the entire energy domain for simulations of the AlTiCrMo test system. This plot is for Method 3 with a 25\% energy sub-domain overlap. The error bars correspond to the standard deviation of time taken of 5 repeat WL simulations with different seeds on the PRNG. These data show that the optimal choices for efficiency are the 1 and 2 walker cases, with the 2 walker case marginally increasing the efficiency of each sub-domain. Additional walkers beyond this appear to confer no further benefit.}
    \label{fig:walkers_speedup}
\end{figure}

Proceeding, we now consider the effect of including more than one walker per energy sub-domain. A natural choice for reducing calculation time in WL is to increase the number of walkers that sample the configuration space, as seen in Ref.~\cite{wl_mpi_2013} where 16 walkers are used per sub-domain. However, evidence of the resulting performance gains has not generally been reported, and we investigate this aspect explicitly here.

Figure~\ref{fig:walkers_speedup} shows a comparison of the speedup per WL instance (taken relative to one walker per energy sub-domain) for different numbers of walkers per energy sub-domain for Method 3, \textit{i.e.} using statically-sized non-uniform energy sub-domains. It can be seen that useful speedup can be obtained, which is across all walker cases. In all cases, the efficiency remains super-linear for the majority if not for all sub-domains, however going beyond around 60 instances the efficiency drops below 100\% while also appearing to plateau in terms of speedup. The initial improvement from the walkers can be attributed to the Central Limit Theorem~\cite{feller1968introduction}. By increasing the number of walkers, the number of samples increases (\textit{e.g.} 2 walkers give twice as many samples, \textit{etc.}) which leads to a reduction in statistical error on the estimate of the DOS proportional to $1/\sqrt{n}$, where $n$ is the number of samples. As the number of walkers increases this reduction in error goes as $1/\sqrt{2}$, $1/\sqrt{3}$, \textit{etc.} which results in a quickly diminishing improvement in efficiency with respect to additional walkers. Our single sub-domain case with a varying quantity of independent walkers can be compared to the implementations found in Refs.~\cite{zierenberg2013, gross2018}, however we only share statistics (\textit{i.e.} the visit histogram and DOS) every 50 sweeps.

\begin{figure}[t]
    \centering
    \includegraphics[width=\linewidth]{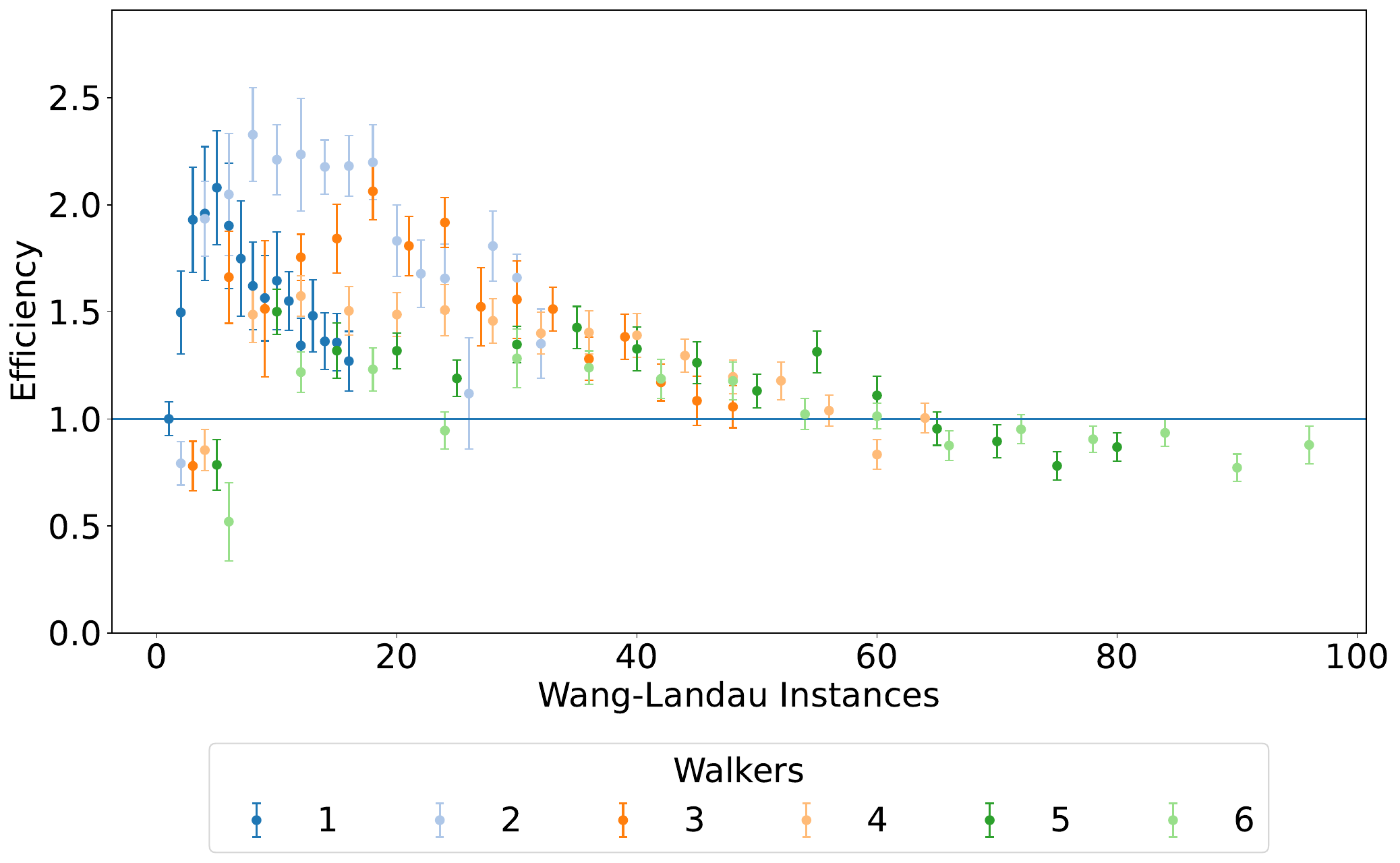}
    \caption{Comparison of the walker efficiency per WL instance for different numbers of walkers per energy sub-domain, taken relative to the case of a single walker across the entire energy domain for simulations of the AlTiCrMo test system. This plot is for Method 3 with a 25\% energy sub-domain overlap. The error bars correspond to the standard deviation of time taken of 5 repeat WL simulations with different seeds on the PRNG. The plot shows the efficiency of each walker introduced across 3 methods, with the 1 and 2 walker cases displaying the highest maximum efficiency per Wang--Landau instance.}
    \label{fig:walker_efficiency}
\end{figure}

We then consider the efficiency of the walkers, as defined in Eq.~\ref{eq:walker_efficiency} above. Figure~\ref{fig:walker_efficiency} shows the efficiency of the instances seen in Figure~\ref{fig:walkers_speedup}. After the introduction of an additional walker, the maximum efficiency exceeds that of the single walker case, however going beyond two walkers per sub-domain, the efficiency begins to decrease substantially which would align with the previously outlined Central Limit Theorem explanation. This increased efficiency from walkers stems partially from the minimum sub-domain size. Since the convergence time is limited by the slowest sub-domain, once the slowest sub-domain reaches the smallest allowed size, further addition of sub-domains will not result in increased efficiency. For the remainder of this study, all results presented are obtained using one walker per energy sub-domain.

\subsubsection{Choice of method for construction of energy sub-domains}

\begin{figure}[t]
    \centering
    \includegraphics[width=\linewidth]{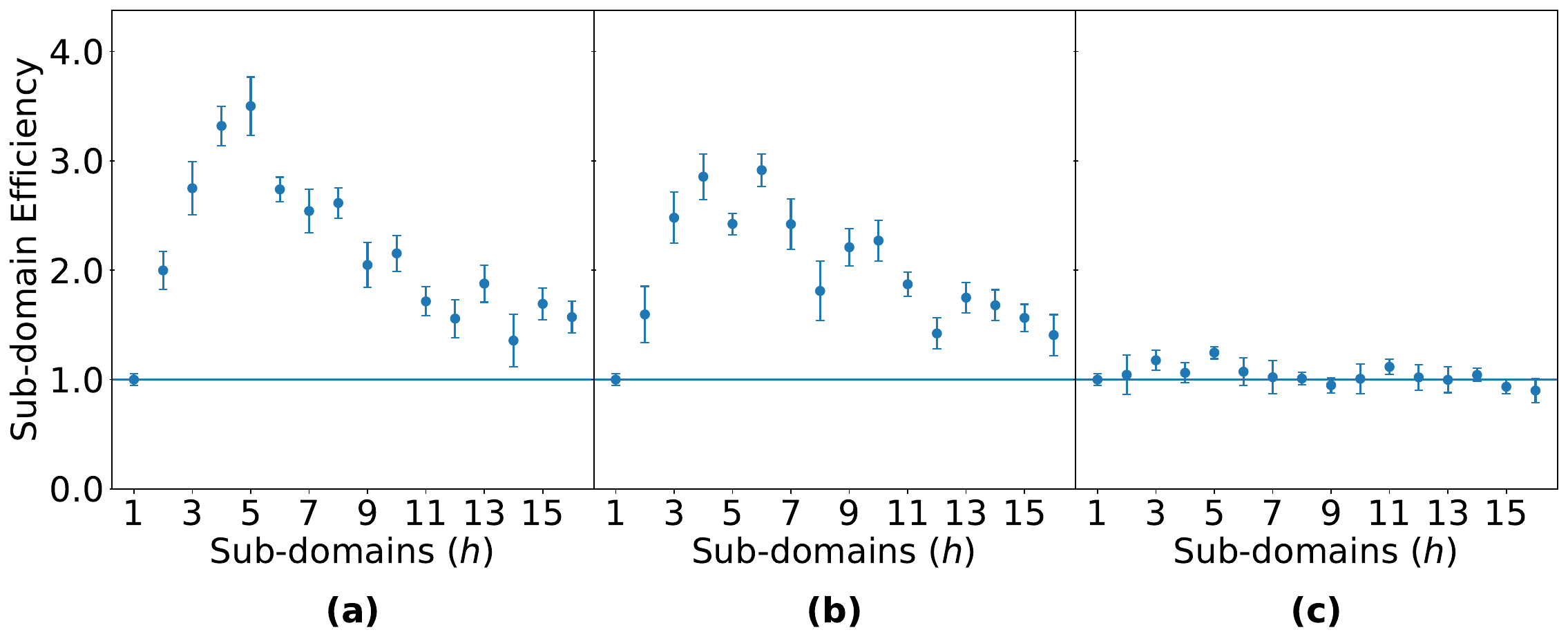}
    \caption{Comparison of the energy sub-domain efficiency relative to the case of a single walker exploring the entire energy domain for simulations of the AlTiCrMo test system. Each method uses one walker per energy sub-domain and a 25\% energy sub-domain overlap, and in all cases replica exchange was enabled. The error bars correspond to the standard deviation of time taken of 5 repeated WL simulations with different seeds on the PRNG. Panel \textbf{(a)} shows Method 1, \textit{i.e.} dynamically load-balanced non-uniform sub-domains. Panel \textbf{(b)} shows Method 3, \textit{i.e.} statically-sized non-uniform sub-domains. Panel \textbf{(c)} shows Method 5, \textit{i.e.} uniformly-sized sub-domains. The plot shows the efficiency of each sub-domain introduced across 3 methods, with Method 1 showing the greatest and longest-lasting efficiency with its efficiency peak occurring at a smaller number of sub-domains. For the case of uniformly-sized energy sub-domains (Method 5), the sub-domain efficiency remains close to, or slightly below, unity regardless of the number of energy sub-domains used.}
    \label{fig:domain_efficiency}
\end{figure}

\begin{figure}[t]
    \centering
    \includegraphics[width=\linewidth]{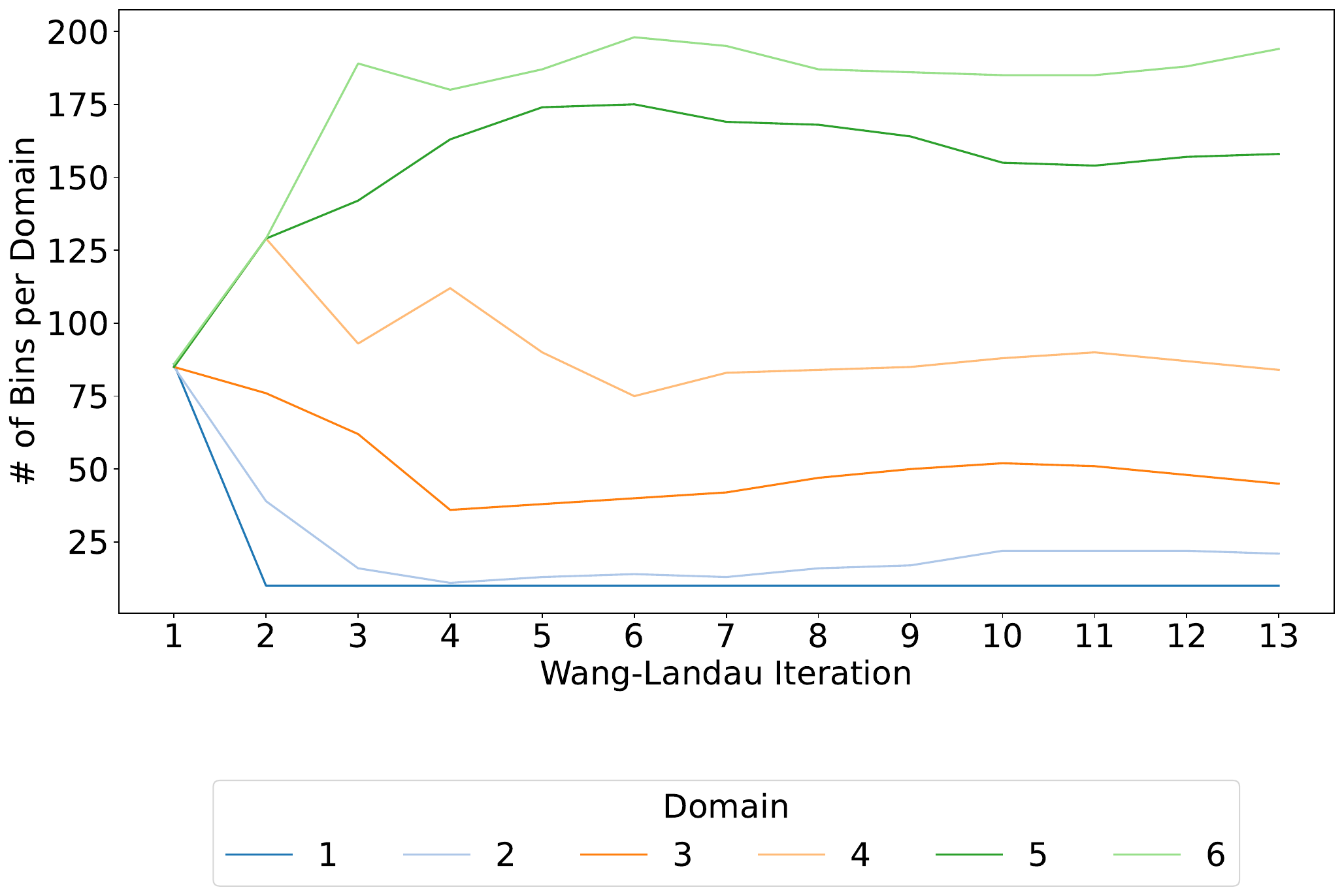}
    \caption{A plot of the relative sizes of energy sub-domains for the proposed dynamic load-balancing scheme as a function of WL iteration number, to demonstrate the effect of load balancing on sub-domain size over the duration of a simulation for simulations of the AlTiCrMo test system. This plot is for Method 1 with one walker per energy sub-domain, a 25\% energy sub-domain overlap, and a decomposition of the total energy domain into 6 sub-domains. The first few iterations show the initial uniform domain decomposition. It can be seen that the load balancing algorithm continues to make substantial adjustments to the relative sizes of energy sub-domains for the first few WL iterations, and that adjustment following an initial pre-sampling step is not enough to achieve good load balancing.}
    \label{fig:load_balance}
\end{figure}

Next, we consider the importance of the use of energy sub-domains. Figure~\ref{fig:domain_efficiency} shows the sub-domain efficiency relative to a single walker operating within the entire energy domain. Within this plot we can see the sub-domain efficiency for Methods 1, 3 and 5 (from left to right respectively) across a range of sub-domains. When the three options for sizing of the energy sub-domains are considered, it can also be seen that the greatest increase in terms of performance is gained by implementing non-uniform energy domain decomposition. By further augmenting the sub-domains and introducing load balancing, a greater efficiency can be maintained at a higher $h$ as compared to using non-uniform energy domain decomposition with static sub-domain sizes. The performance gain observed from energy domain decomposition can be attributed to a reduction in the time taken for any walker to diffuse across the sub-domain, which reduces wasted sampling time spent on easily converged bins. Further discussion of this aspect can be found in Sec.~\ref{sec:super-linear}. 

To demonstrate the efficacy of our proposed dynamic energy sub-domain sizing scheme, we plot the comparative sizes of the energy sub-domains for an illustrative simulation run in Figure~\ref{fig:load_balance}, for the case where a total of six energy sub-domains were used. The first iteration here corresponds to the pre-sampling step with initial uniform domain decomposition. It can be seen that there is a difference in the size of the sub-domains at iteration 2 as compared to iteration 13. This demonstrates the difficulty of \textit{a priori} load balancing in the pre-sampling step and the benefits conferred by iterative load balancing. Also of note is the fact that the relative sizes of domains settle over the course of many iterations, demonstrating the numerical stability of our proposed load-balancing scheme.

\begin{figure*}[t]
    \centering
    \includegraphics[width=\linewidth]{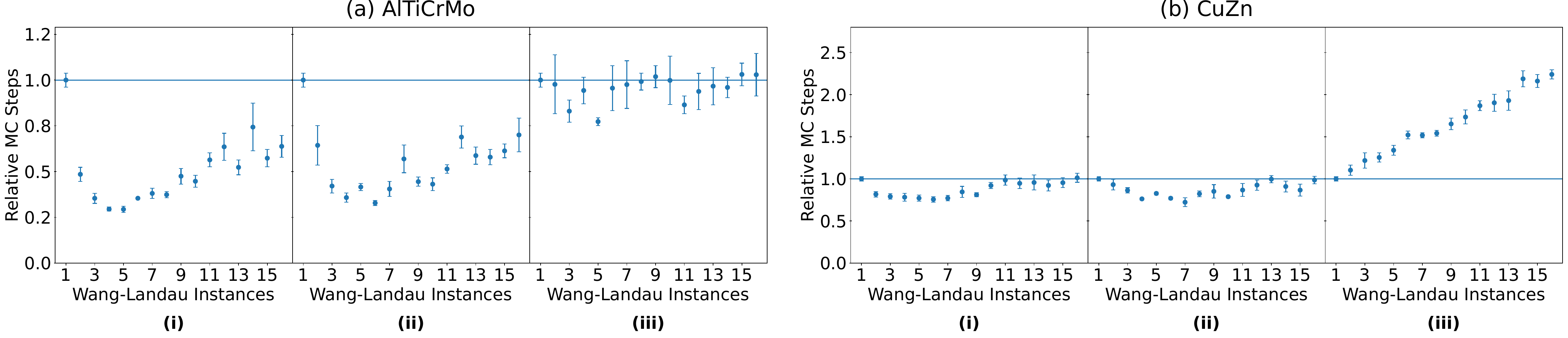}
    \caption{MC steps taken to complete all WL sampling iterations relative to a single walker across the entire energy domain for AlTiCrMo and CuZn. Each method uses one walker and 25\% energy sub-domain overlap. Panel (i) shows results for Method 1, \textit{i.e.} dynamically load-balanced non-uniform sub-domains. Panel (ii) shows results for Method 3, \textit{i.e.} statically-sized non-uniform sub-domains. Finally, panel (iii) shows results for Method 5, \textit{i.e.} uniformly-sized sub-domains. Both plots show that domain decomposition reduces the overall number of MC steps required to converge the WL sampling simulation.}
    \label{fig:mc_steps}
\end{figure*}

\begin{figure*}[t]
    \centering
    \includegraphics[width=\linewidth]{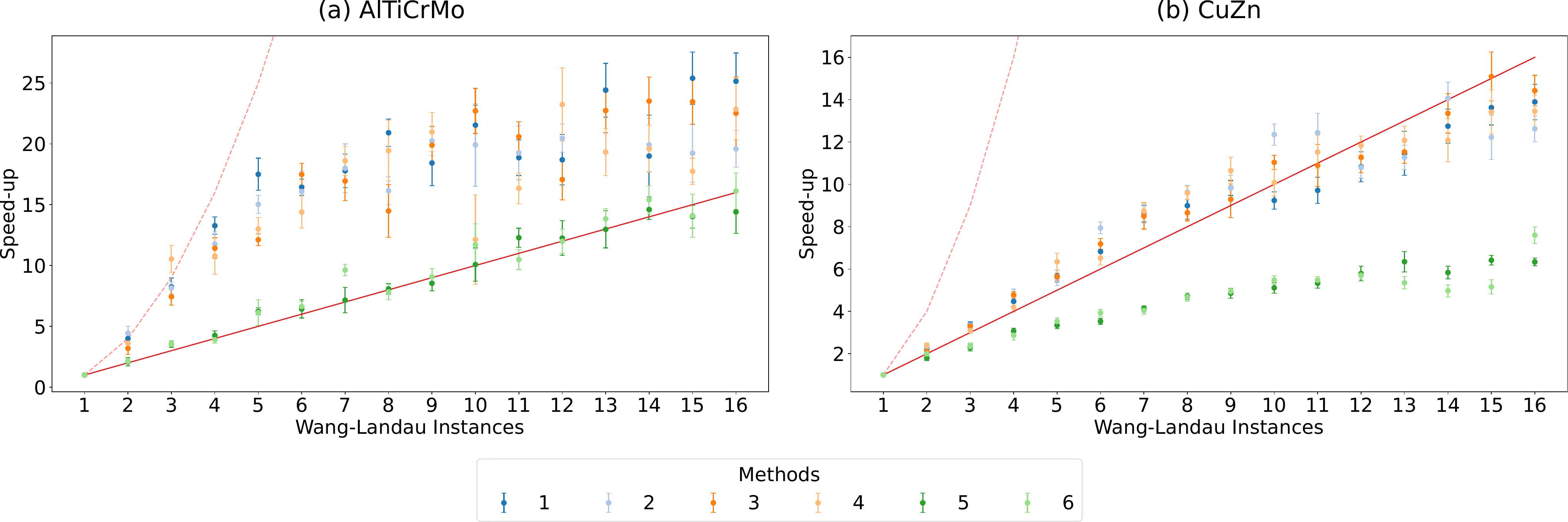}
    \caption{Comparison of the speedup per WL instance, taken relative to the case of a single walker across the entire energy domain, for simulations of AlTiCrMo and CuZn. The dotted line shows sub-domains squared $(h^2)$. Each method has one walker and 25\% overlap. Both plots show the greatest increase in performance can be obtained through non-uniform domain decomposition with a modest further increase from load balancing.}
    \label{fig:overall_speedup}
\end{figure*}

\subsubsection{Origins of observed super-linear efficiency}
\label{sec:super-linear}

Now that greater than 100\% efficiency for energy sub-domain decomposition has been demonstrated (evidenced in Figures~\ref{fig:overlap_nrmse_speedup} and \ref{fig:walkers_speedup}), we support our initial analysis given in Sec.~\ref{sec:performance_metrics} and confirm that super-linear efficiency results in a reduction in the total number of MC steps needed (summed over all WL instances) to converge the DOS. This is broadly equivalent to the statement that division of the total energy domain into sub-domains---and restricting walkers to operate in these sub-domains---reduces the total number of MC steps required to adequately explore the configuration space for the construction of the DOS.

Figure~\ref{fig:mc_steps} shows the relative total number of MC steps required to complete all WL sampling iterations as a function of WL instance count, \textit{i.e.} number of energy sub-domains, as compared to the case of a single walker operating across the entire energy domain. For the AlTiCrMo case shown in panel (a), it can be seen that the total number of MC steps required to converge the simulation substantially drops once energy sub-domains are introduced. For the case of uniform energy sub-domains shown in panel (a.iii), as the number of sub-domains increases the total number of MC steps (sum of all steps taken across all $m\cdot h$ walkers) increases. In our conceptual model of a diffusive random walk in energy, this lack of reduction in MC steps required is due to the non-uniformity of `diffusivity', $D$, quantifying the speed of walker exploration across the energy domain. In turn, this results in the WL convergence being dominated by the sub-domain which has the lowest $D$. However, for the two cases in which non-uniform energy sub-domains are employed (panels (a.i) and (a.ii)), the total number of MC steps needed to fully converge the simulation decreases, which is particularly prominent for the first few additional WL instances, \textit{i.e.} energy sub-domains. In this case, not only is the total work across the entire energy domain divided across WL instances, the total amount of work to be done is reduced (by around a factor of two for the case of two WL instances, \textit{i.e.} energy sub-domains), which broadly agrees with the $\sim 4 \times$ speedup seen in particular towards the left-hand data points of panel (a) in Figure~\ref{fig:overall_speedup}. This analysis confirms our earlier assertion that the super-linear speedup, observed in both this study and in an earlier work, can be concretely understood in terms of a reduction in the total number of MC steps, \textit{i.e.} the work, required to converge a given simulation.

\subsubsection{AlTiCrMo results summary}
\label{sec:summary}

To summarise the key findings of our study, in Figure~\ref{fig:overall_speedup} we compare the overall simulation speedup (the most relevant metric for parallel performance of an algorithm for applications) across all six of the studied methods, \textit{i.e.} parallelisation schemes. Here, all data points presented are for simulations run using a 25\% energy sub-domain overlap and one walker per energy sub-domain. 

Focusing on plot (a) of Figure~\ref{fig:overall_speedup}, it can be seen that all of the proposed methods achieve an appreciable speedup, with even the worst-performing methods (5 and 6) achieving near linear speedup out to 16 WL instances, \textit{i.e.} energy sub-domains. However, the methods which stand out are 1, 2, 3, and 4, \textit{i.e.} those using non-uniform energy domain decomposition. This result reinforces the idea that the greatest gain in performance is obtained via non-uniform domain decomposition. Methods 1 and 2, which make use of the proposed dynamic energy sub-domain sizing scheme, offer modest improvements over the case of statically-sized non-uniform energy sub-domains. The increase in performance between the load-balanced (Methods 1, 2) and not load balanced (Methods 3, 4) allows for the simulation to retain its sub-domain efficiency for a greater number of sub-domains giving access to greater speedups. Of note, replica exchange does not appear to significantly impact performance either positively or negatively.

\subsection{CuZn}

Now that the effects of the parallelisation schemes have been outlined for AlTiCrMo, we will contrast the results to the more simple CuZn alloy. For this benchmark, we use a simulation supercell made up of $8 \times 8 \times 8$ bcc cubic unit cells for a total of 1024 atoms. We find that 1024 atoms is necessary for an appreciable number of energy bins for the purposes of load balancing due to the degenerate energy landscape found within this binary alloy. When running the same number of atoms as with the AlTiCrMo case (128 atoms) the simulation only requires 16 bins and takes $\sim$2.5 seconds to complete 12 WL iterations (one WL instance across the entire energy domain), further demonstrating the challenging nature of the AlTiCrMo system. Parallelisation is, however, still useful in the simpler CuZn case when simulating a larger system. Our WL sampling simulations used $N=256$ energy bins over an energy domain of $E=-76$~meV/atom to $E=-5$~meV/atom, chosen via the method outlined in Sec.~\ref{sec:wl_sampling} with $T_{\text{min}}=50K$ and $T_{\text{max}}=3000K$.

For CuZn, we select the two most informative metrics for comparison: the relative MC steps and total speedup. These results can be seen in plots (b) of Figures~\ref{fig:mc_steps} and \ref{fig:overall_speedup} respectively. Starting with plot (b) in Figure~\ref{fig:mc_steps}, it can be seen that the total number of MC steps required to converge the simulation drops once non-uniform energy sub-domains are introduced. This effect is less pronounced than the AlTiCrMo case, however panels (b.i) and (b.ii) display a similar trend of an initial reduction in MC steps required before subsequently increasing. Panel (b.iii) is of particular interest, as it exhibits an evidently different behaviour, with increasing the number of uniform domains directly increasing the total number of MC steps required to converge the simulation. This increase in the required work, which would also apply to the methods in panels (b.i) and (b.ii), is caused by the difficulty of converging low energy bins. Using the case in panel (b.iii), as more sub-domains are introduced their size decreases, however, the difficult to converge sub-domain never becomes small enough to effectively divide the load. This issue appears in cases (b.i) and (b.ii) due to the limit imposed on the minimum size of a sub-domain described in Section~\ref{sec:load_balance}. In both cases this results in an overall increase in the number of MC steps, as each iteration requires the low energy domain to converge before completing.

Proceeding, we now look to Figure~\ref{fig:overall_speedup} and begin by addressing the difference in magnitude of speedup between AlTiCrMo and CuZn. CuZn exhibits a decreased overall benefit from parallelisation, which is most evident by the speedup not reaching the theoretical $h^2$ threshold. This can be attributed to the discretised DOS and degenerate energy states (as compared to AlTiCrMo) which result in particularly non-uniform convergence times across the energy domain. Both plots also exhibit a similar trend between methods that have and do not have non-uniform domain decomposition, which shows that the key to unlocking higher parallelisation is the introduction of non-uniform domains (particularly evident in the CuZn case).

\section{Summary, Conclusions, and Recommendations}
\label{sec:conclusions_and_recommendations}

In summary, several parallelisation strategies for the WL sampling algorithm have been presented, implemented, and their performance assessed. The schemes studied, both in isolation and in combination, included energy domain decomposition, replica exchange, and a dynamic load-balancing scheme proposed in this work. From the performance evaluation within this study, several important insights emerge for the effective implementation of parallel WL sampling, which should---in principle---be extensible to other flat-histogram methods such as transition matrix MC~\cite{wang_transition_1999}. 

The most significant factor in determining parallel performance is the choice of energy domain decomposition scheme, with use of non-uniform energy sub-domains consistently providing the largest gains in both speedup and overall simulation efficiency. Dynamic load balancing offers a modest further enhancement in performance by allowing the energy domain decomposition scheme to adjust itself `on the fly' and adapt to the sampling demands of complex energy landscapes, with energy sub-domain boundaries shifting between each WL iteration. Careful attention to the number of walkers allocated to each energy sub-domain also proves critical. While employing multiple walkers may seem like a natural choice when seeking to parallelise the WL sampling algorithm, our results indicate that this produces rapidly diminishing returns as a function of number of walkers, and potentially a reduction in efficiency. In this study, we found that the use of one or two walkers per sub-domain is generally sufficient to realise any performance improvements available through the use of multiple walkers per energy sub-domain. Also of note is that replica exchange does not noticeably influence efficiency in this test case, but nor does it compromise performance, which suggests that it may be included where needed to improve sampling without penalty. This is likely due to the non-physical MC moves (long-range atomic swaps) in our test system which result in the ergodicity not being limited by the need to cross energy barriers. Finally, the construction of overlap regions between sub-domains is shown to have a negligible effect on performance unless it is in excess of a 75\% overlap, underscoring that large overlap is necessary only for replica exchange and not for the reliable combination of local DOS histograms to form the global DOS. We would also like to highlight that the biggest overall reductions in the total MC steps needed to converge the WL sampling simulations are more prominent for the more challenging of the two benchmark systems (AlTiCrMo) and we expect the largest benefit from the parallelisation schemes to be found in such scenarios, \textit{i.e.} complex systems with quasi-continuous energy landscapes.

\subsection{Recommendations}

On the basis of these findings, we recommend that parallel implementations of WL sampling prioritise using non-uniform energy domain decomposition as a primary route to improved simulation efficiency. Subsequently, one should implement multiple walkers per sub-domain for further efficiency and performance. Once those two features have been implemented, dynamic load balancing should subsequently be incorporated wherever feasible for modest improvement of per-instance efficiency. If reduced runtime is more important than efficiency, and if adding further sub-domains hurts performance, further increasing the number of walkers can be employed as a complementary tool, though they will likely decrease overall computational efficiency. Finally, overlap regions of a modest size should be included primarily to facilitate replica exchange, as they have little to no impact on the accuracy of the reconstructed total DOS, though excessively-sized overlap regions should be avoided due to the associated drop in parallel performance.

\section*{CRediT authorship contribution statement}
\textbf{Hubert J. Naguszewski} Conceptualization (equal), Data curation, Investigation (lead), Methodology (lead), Software (lead), Validation, Visualization, Writing - original draft (lead), Writing - review and editing (equal). \textbf{Christopher D. Woodgate} Investigation (supporting), Methodology (supporting), Software (supporting), Supervision (supporting), Writing – original draft (supporting), Writing – review and editing (equal). \textbf{David Quigley} Conceptualization (equal), Funding acquisition, Project administration, Resources, Supervision (lead), Writing – review and editing (equal).

\section*{Declaration of competing interest}
The authors declare that they have no known competing financial interests or personal relationships that could have appeared to influence the work reported in this paper.

\section*{Data Availability}
The data supporting the findings of this study are freely available at the following DOI/URL: \href{https://doi.org/10.5281/zenodo.18921874}{https://doi.org/10.5281/zenodo.18921874}. The \textsc{Brawl} package, which was used to obtain all results presented in this work is fully open-source and available at \href{https://github.com/ChrisWoodgate/BraWl}{https://github.com/ChrisWoodgate/BraWl}.

\begin{acknowledgments}
H.J.N. acknowledges a funded studentship within the UK Engineering and Physical Sciences Research Council (EPSRC) Centre for Doctoral Training in Modelling of Heterogeneous Systems, Grant EP/S022848/1. C.D.W. acknowledges support from an EPSRC Doctoral Prize Fellowship at the University of Bristol, Grant EP/W524414/1. Calculations were performed using the Sulis Tier 2 HPC platform hosted by the Scientific Computing Research Technology Platform (SCRTP) at the University of Warwick. Sulis is funded by EPSRC Grant EP/T022108/1 and the HPC Midlands+ consortium. Additional computing facilities were provided by the Scientific Computing Research Technology Platform at the University of Warwick, and by the Advanced Computing Research Centre at the University of Bristol.
\end{acknowledgments}


\begin{thebibliography}{76}%
\makeatletter
\providecommand \@ifxundefined [1]{%
 \@ifx{#1\undefined}
}%
\providecommand \@ifnum [1]{%
 \ifnum #1\expandafter \@firstoftwo
 \else \expandafter \@secondoftwo
 \fi
}%
\providecommand \@ifx [1]{%
 \ifx #1\expandafter \@firstoftwo
 \else \expandafter \@secondoftwo
 \fi
}%
\providecommand \natexlab [1]{#1}%
\providecommand \enquote  [1]{``#1''}%
\providecommand \bibnamefont  [1]{#1}%
\providecommand \bibfnamefont [1]{#1}%
\providecommand \citenamefont [1]{#1}%
\providecommand \href@noop [0]{\@secondoftwo}%
\providecommand \href [0]{\begingroup \@sanitize@url \@href}%
\providecommand \@href[1]{\@@startlink{#1}\@@href}%
\providecommand \@@href[1]{\endgroup#1\@@endlink}%
\providecommand \@sanitize@url [0]{\catcode `\\12\catcode `\$12\catcode `\&12\catcode `\#12\catcode `\^12\catcode `\_12\catcode `\%12\relax}%
\providecommand \@@startlink[1]{}%
\providecommand \@@endlink[0]{}%
\providecommand \url  [0]{\begingroup\@sanitize@url \@url }%
\providecommand \@url [1]{\endgroup\@href {#1}{\urlprefix }}%
\providecommand \urlprefix  [0]{URL }%
\providecommand \Eprint [0]{\href }%
\providecommand \doibase [0]{http://dx.doi.org/}%
\providecommand \selectlanguage [0]{\@gobble}%
\providecommand \bibinfo  [0]{\@secondoftwo}%
\providecommand \bibfield  [0]{\@secondoftwo}%
\providecommand \translation [1]{[#1]}%
\providecommand \BibitemOpen [0]{}%
\providecommand \bibitemStop [0]{}%
\providecommand \bibitemNoStop [0]{.\EOS\space}%
\providecommand \EOS [0]{\spacefactor3000\relax}%
\providecommand \BibitemShut  [1]{\csname bibitem#1\endcsname}%
\let\auto@bib@innerbib\@empty
\bibitem [{\citenamefont {Landau}\ and\ \citenamefont {Binder}(2021)}]{landau_guide_2021}%
  \BibitemOpen
  \bibfield  {author} {\bibinfo {author} {\bibfnamefont {D.~P.}\ \bibnamefont {Landau}}\ and\ \bibinfo {author} {\bibfnamefont {K.}~\bibnamefont {Binder}},\ }\href {\doibase 10.1017/9781108780346} {\emph {\bibinfo {title} {A {Guide} to {Monte} {Carlo} {Simulations} in {Statistical} {Physics}}}},\ \bibinfo {edition} {5th}\ ed.\ (\bibinfo  {publisher} {Cambridge University Press},\ \bibinfo {address} {Cambridge, UK},\ \bibinfo {year} {2021})\BibitemShut {NoStop}%
\bibitem [{\citenamefont {Metropolis}\ \emph {et~al.}(1953)\citenamefont {Metropolis}, \citenamefont {Rosenbluth}, \citenamefont {Rosenbluth}, \citenamefont {Teller},\ and\ \citenamefont {Teller}}]{metropolis_equation_1953}%
  \BibitemOpen
  \bibfield  {author} {\bibinfo {author} {\bibfnamefont {N.}~\bibnamefont {Metropolis}}, \bibinfo {author} {\bibfnamefont {A.~W.}\ \bibnamefont {Rosenbluth}}, \bibinfo {author} {\bibfnamefont {M.~N.}\ \bibnamefont {Rosenbluth}}, \bibinfo {author} {\bibfnamefont {A.~H.}\ \bibnamefont {Teller}}, \ and\ \bibinfo {author} {\bibfnamefont {E.}~\bibnamefont {Teller}},\ }\bibfield  {title} {\enquote {\bibinfo {title} {Equation of {State} {Calculations} by {Fast} {Computing} {Machines}},}\ }\href {\doibase 10.1063/1.1699114} {\bibfield  {journal} {\bibinfo  {journal} {The Journal of Chemical Physics}\ }\textbf {\bibinfo {volume} {21}},\ \bibinfo {pages} {1087} (\bibinfo {year} {1953})}\BibitemShut {NoStop}%
\bibitem [{\citenamefont {Frenkel}\ and\ \citenamefont {Smit}(2002)}]{frenkel_understanding_2002}%
  \BibitemOpen
  \bibfield  {author} {\bibinfo {author} {\bibfnamefont {D.}~\bibnamefont {Frenkel}}\ and\ \bibinfo {author} {\bibfnamefont {B.}~\bibnamefont {Smit}},\ }\href@noop {} {\emph {\bibinfo {title} {Understanding molecular simulation: from algorithms to applications}}},\ \bibinfo {edition} {2nd}\ ed.,\ \bibinfo {series} {Computational science series}\ No.~\bibinfo {number} {1}\ (\bibinfo  {publisher} {Academic Press},\ \bibinfo {year} {2002})\BibitemShut {NoStop}%
\bibitem [{\citenamefont {Bennett}(1976)}]{bennett_efficient_1976}%
  \BibitemOpen
  \bibfield  {author} {\bibinfo {author} {\bibfnamefont {C.~H.}\ \bibnamefont {Bennett}},\ }\bibfield  {title} {\enquote {\bibinfo {title} {Efficient estimation of free energy differences from {Monte} {Carlo} data},}\ }\href {\doibase 10.1016/0021-9991(76)90078-4} {\bibfield  {journal} {\bibinfo  {journal} {Journal of Computational Physics}\ }\textbf {\bibinfo {volume} {22}},\ \bibinfo {pages} {245} (\bibinfo {year} {1976})}\BibitemShut {NoStop}%
\bibitem [{\citenamefont {Torrie}\ and\ \citenamefont {Valleau}(1977)}]{torrie_nonphysical_1977}%
  \BibitemOpen
  \bibfield  {author} {\bibinfo {author} {\bibfnamefont {G.~M.}\ \bibnamefont {Torrie}}\ and\ \bibinfo {author} {\bibfnamefont {J.~P.}\ \bibnamefont {Valleau}},\ }\bibfield  {title} {\enquote {\bibinfo {title} {Nonphysical sampling distributions in {Monte} {Carlo} free-energy estimation: {Umbrella} sampling},}\ }\href {\doibase 10.1016/0021-9991(77)90121-8} {\bibfield  {journal} {\bibinfo  {journal} {Journal of Computational Physics}\ }\textbf {\bibinfo {volume} {23}},\ \bibinfo {pages} {187} (\bibinfo {year} {1977})}\BibitemShut {NoStop}%
\bibitem [{\citenamefont {Berg}\ and\ \citenamefont {Neuhaus}(1991)}]{berg_multicanonical_1991}%
  \BibitemOpen
  \bibfield  {author} {\bibinfo {author} {\bibfnamefont {B.~A.}\ \bibnamefont {Berg}}\ and\ \bibinfo {author} {\bibfnamefont {T.}~\bibnamefont {Neuhaus}},\ }\bibfield  {title} {\enquote {\bibinfo {title} {Multicanonical algorithms for first order phase transitions},}\ }\href {\doibase 10.1016/0370-2693(91)91256-U} {\bibfield  {journal} {\bibinfo  {journal} {Physics Letters B}\ }\textbf {\bibinfo {volume} {267}},\ \bibinfo {pages} {249} (\bibinfo {year} {1991})}\BibitemShut {NoStop}%
\bibitem [{\citenamefont {Berg}\ and\ \citenamefont {Neuhaus}(1992)}]{berg_multicanonical_1992}%
  \BibitemOpen
  \bibfield  {author} {\bibinfo {author} {\bibfnamefont {B.~A.}\ \bibnamefont {Berg}}\ and\ \bibinfo {author} {\bibfnamefont {T.}~\bibnamefont {Neuhaus}},\ }\bibfield  {title} {\enquote {\bibinfo {title} {Multicanonical ensemble: {A} new approach to simulate first-order phase transitions},}\ }\href {\doibase 10.1103/PhysRevLett.68.9} {\bibfield  {journal} {\bibinfo  {journal} {Physical Review Letters}\ }\textbf {\bibinfo {volume} {68}},\ \bibinfo {pages} {9} (\bibinfo {year} {1992})}\BibitemShut {NoStop}%
\bibitem [{\citenamefont {Lee}(1993)}]{lee_new_1993}%
  \BibitemOpen
  \bibfield  {author} {\bibinfo {author} {\bibfnamefont {J.}~\bibnamefont {Lee}},\ }\bibfield  {title} {\enquote {\bibinfo {title} {New {Monte} {Carlo} algorithm: {Entropic} sampling},}\ }\href {\doibase 10.1103/PhysRevLett.71.211} {\bibfield  {journal} {\bibinfo  {journal} {Physical Review Letters}\ }\textbf {\bibinfo {volume} {71}},\ \bibinfo {pages} {211} (\bibinfo {year} {1993})}\BibitemShut {NoStop}%
\bibitem [{\citenamefont {Wang}(1999)}]{wang_transition_1999}%
  \BibitemOpen
  \bibfield  {author} {\bibinfo {author} {\bibfnamefont {J.-S.}\ \bibnamefont {Wang}},\ }\bibfield  {title} {\enquote {\bibinfo {title} {Transition matrix {Monte} {Carlo} method},}\ }\href {\doibase https://doi.org/10.1016/S0010-4655(99)00270-2} {\bibfield  {journal} {\bibinfo  {journal} {Computer Physics Communications}\ }\textbf {\bibinfo {volume} {121-122}},\ \bibinfo {pages} {22} (\bibinfo {year} {1999})},\ \bibinfo {note} {proceedings of the Europhysics Conference on Computational Physics CCP 1998}\BibitemShut {NoStop}%
\bibitem [{\citenamefont {Wang}\ and\ \citenamefont {Landau}(2001{\natexlab{a}})}]{wang_efficient_2001}%
  \BibitemOpen
  \bibfield  {author} {\bibinfo {author} {\bibfnamefont {F.}~\bibnamefont {Wang}}\ and\ \bibinfo {author} {\bibfnamefont {D.~P.}\ \bibnamefont {Landau}},\ }\bibfield  {title} {\enquote {\bibinfo {title} {Efficient, {Multiple}-{Range} {Random} {Walk} {Algorithm} to {Calculate} the {Density} of {States}},}\ }\href {\doibase 10.1103/PhysRevLett.86.2050} {\bibfield  {journal} {\bibinfo  {journal} {Physical Review Letters}\ }\textbf {\bibinfo {volume} {86}},\ \bibinfo {pages} {2050} (\bibinfo {year} {2001}{\natexlab{a}})}\BibitemShut {NoStop}%
\bibitem [{\citenamefont {Wang}\ and\ \citenamefont {Landau}(2001{\natexlab{b}})}]{wang_determining_2001}%
  \BibitemOpen
  \bibfield  {author} {\bibinfo {author} {\bibfnamefont {F.}~\bibnamefont {Wang}}\ and\ \bibinfo {author} {\bibfnamefont {D.~P.}\ \bibnamefont {Landau}},\ }\bibfield  {title} {\enquote {\bibinfo {title} {Determining the density of states for classical statistical models: {A} random walk algorithm to produce a flat histogram},}\ }\href {\doibase 10.1103/PhysRevE.64.056101} {\bibfield  {journal} {\bibinfo  {journal} {Physical Review E}\ }\textbf {\bibinfo {volume} {64}},\ \bibinfo {pages} {056101} (\bibinfo {year} {2001}{\natexlab{b}})}\BibitemShut {NoStop}%
\bibitem [{\citenamefont {Landau}\ \emph {et~al.}(2004)\citenamefont {Landau}, \citenamefont {Tsai},\ and\ \citenamefont {Exler}}]{landau_new_2004}%
  \BibitemOpen
  \bibfield  {author} {\bibinfo {author} {\bibfnamefont {D.~P.}\ \bibnamefont {Landau}}, \bibinfo {author} {\bibfnamefont {S.-H.}\ \bibnamefont {Tsai}}, \ and\ \bibinfo {author} {\bibfnamefont {M.}~\bibnamefont {Exler}},\ }\bibfield  {title} {\enquote {\bibinfo {title} {A new approach to {Monte} {Carlo} simulations in statistical physics: {Wang}-{Landau} sampling},}\ }\href {\doibase 10.1119/1.1707017} {\bibfield  {journal} {\bibinfo  {journal} {American Journal of Physics}\ }\textbf {\bibinfo {volume} {72}},\ \bibinfo {pages} {1294} (\bibinfo {year} {2004})}\BibitemShut {NoStop}%
\bibitem [{\citenamefont {Landau}\ and\ \citenamefont {Wang}(2002)}]{landau_determining_2002}%
  \BibitemOpen
  \bibfield  {author} {\bibinfo {author} {\bibfnamefont {D.~P.}\ \bibnamefont {Landau}}\ and\ \bibinfo {author} {\bibfnamefont {F.}~\bibnamefont {Wang}},\ }\bibfield  {title} {\enquote {\bibinfo {title} {Determining the density of states for classical statistical models by a flat-histogram random walk},}\ }\href {\doibase 10.1016/S0010-4655(02)00374-0} {\bibfield  {journal} {\bibinfo  {journal} {Computer Physics Communications}\ }\bibinfo {series} {Proceedings of the {Europhysics} {Conference} on {Computational} {Physics} {Computational} {Modeling} and {Simulation} of {Complex} {Systems}},\ \textbf {\bibinfo {volume} {147}},\ \bibinfo {pages} {674} (\bibinfo {year} {2002})}\BibitemShut {NoStop}%
\bibitem [{\citenamefont {Brown}\ \emph {et~al.}(2011)\citenamefont {Brown}, \citenamefont {Rusanu}, \citenamefont {Daene}, \citenamefont {Nicholson}, \citenamefont {Eisenbach},\ and\ \citenamefont {Fidler}}]{brown_improved_2011}%
  \BibitemOpen
  \bibfield  {author} {\bibinfo {author} {\bibfnamefont {G.}~\bibnamefont {Brown}}, \bibinfo {author} {\bibfnamefont {A.}~\bibnamefont {Rusanu}}, \bibinfo {author} {\bibfnamefont {M.}~\bibnamefont {Daene}}, \bibinfo {author} {\bibfnamefont {D.~M.}\ \bibnamefont {Nicholson}}, \bibinfo {author} {\bibfnamefont {M.}~\bibnamefont {Eisenbach}}, \ and\ \bibinfo {author} {\bibfnamefont {J.}~\bibnamefont {Fidler}},\ }\bibfield  {title} {\enquote {\bibinfo {title} {Improved methods for calculating thermodynamic properties of magnetic systems using {Wang}-{Landau} density of states},}\ }\href {\doibase 10.1063/1.3565413} {\bibfield  {journal} {\bibinfo  {journal} {Journal of Applied Physics}\ }\textbf {\bibinfo {volume} {109}},\ \bibinfo {pages} {07E161} (\bibinfo {year} {2011})}\BibitemShut {NoStop}%
\bibitem [{\citenamefont {Volkov}\ \emph {et~al.}(2012)\citenamefont {Volkov}, \citenamefont {Lyubartsev},\ and\ \citenamefont {Bergström}}]{volkov_phase_2012}%
  \BibitemOpen
  \bibfield  {author} {\bibinfo {author} {\bibfnamefont {N.}~\bibnamefont {Volkov}}, \bibinfo {author} {\bibfnamefont {A.}~\bibnamefont {Lyubartsev}}, \ and\ \bibinfo {author} {\bibfnamefont {L.}~\bibnamefont {Bergström}},\ }\bibfield  {title} {\enquote {\bibinfo {title} {Phase transitions and thermodynamic properties of dense assemblies of truncated nanocubes and cuboctahedra},}\ }\href {\doibase 10.1039/C2NR30411B} {\bibfield  {journal} {\bibinfo  {journal} {Nanoscale}\ }\textbf {\bibinfo {volume} {4}},\ \bibinfo {pages} {4765} (\bibinfo {year} {2012})}\BibitemShut {NoStop}%
\bibitem [{\citenamefont {Lee}\ \emph {et~al.}(2024)\citenamefont {Lee}, \citenamefont {Kim},\ and\ \citenamefont {Kim}}]{lee_frustrated_2024}%
  \BibitemOpen
  \bibfield  {author} {\bibinfo {author} {\bibfnamefont {J.~H.}\ \bibnamefont {Lee}}, \bibinfo {author} {\bibfnamefont {S.-Y.}\ \bibnamefont {Kim}}, \ and\ \bibinfo {author} {\bibfnamefont {J.~M.}\ \bibnamefont {Kim}},\ }\bibfield  {title} {\enquote {\bibinfo {title} {Frustrated {Ising} model with competing interactions on a square lattice},}\ }\href {\doibase 10.1103/PhysRevB.109.064422} {\bibfield  {journal} {\bibinfo  {journal} {Physical Review B}\ }\textbf {\bibinfo {volume} {109}},\ \bibinfo {pages} {064422} (\bibinfo {year} {2024})}\BibitemShut {NoStop}%
\bibitem [{\citenamefont {Sato}\ \emph {et~al.}(2010)\citenamefont {Sato}, \citenamefont {Takizawa},\ and\ \citenamefont {Mohri}}]{sato_wanglandau_2010}%
  \BibitemOpen
  \bibfield  {author} {\bibinfo {author} {\bibfnamefont {K.}~\bibnamefont {Sato}}, \bibinfo {author} {\bibfnamefont {S.}~\bibnamefont {Takizawa}}, \ and\ \bibinfo {author} {\bibfnamefont {T.}~\bibnamefont {Mohri}},\ }\bibfield  {title} {\enquote {\bibinfo {title} {A {Wang}–{Landau} {Monte} {Carlo} {Simulation} of {Melting} in fcc {Lennard}-{Jones} {System}},}\ }\href {\doibase 10.1143/JPSJ.79.084602} {\bibfield  {journal} {\bibinfo  {journal} {Journal of the Physical Society of Japan}\ }\textbf {\bibinfo {volume} {79}},\ \bibinfo {pages} {084602} (\bibinfo {year} {2010})}\BibitemShut {NoStop}%
\bibitem [{\citenamefont {Pei}\ \emph {et~al.}()\citenamefont {Pei}, \citenamefont {Eisenbach}, \citenamefont {Mu},\ and\ \citenamefont {Stocks}}]{pei_error_2019}%
  \BibitemOpen
  \bibfield  {author} {\bibinfo {author} {\bibfnamefont {Z.}~\bibnamefont {Pei}}, \bibinfo {author} {\bibfnamefont {M.}~\bibnamefont {Eisenbach}}, \bibinfo {author} {\bibfnamefont {S.}~\bibnamefont {Mu}}, \ and\ \bibinfo {author} {\bibfnamefont {G.~M.}\ \bibnamefont {Stocks}},\ }\bibfield  {title} {\enquote {\bibinfo {title} {Error controlling of the combined {Cluster}-{Expansion} and {Wang}–{Landau} {Monte}-{Carlo} method and its application to {FeCo}},}\ }\href {\doibase 10.1016/j.cpc.2018.09.017} {\bibfield  {journal} {\bibinfo  {journal} {Computer Physics Communications}\ }\textbf {\bibinfo {volume} {235}},\ \bibinfo {pages} {95}}\BibitemShut {NoStop}%
\bibitem [{\citenamefont {Wang}\ and\ \citenamefont {He}(2011)}]{wang_phase_2011}%
  \BibitemOpen
  \bibfield  {author} {\bibinfo {author} {\bibfnamefont {Z.}~\bibnamefont {Wang}}\ and\ \bibinfo {author} {\bibfnamefont {X.}~\bibnamefont {He}},\ }\bibfield  {title} {\enquote {\bibinfo {title} {Phase transition of a single star polymer: a {Wang}-{Landau} sampling study},}\ }\href {\doibase 10.1063/1.3629849} {\bibfield  {journal} {\bibinfo  {journal} {The Journal of Chemical Physics}\ }\textbf {\bibinfo {volume} {135}},\ \bibinfo {pages} {094902} (\bibinfo {year} {2011})}\BibitemShut {NoStop}%
\bibitem [{\citenamefont {Antypov}\ and\ \citenamefont {Elliott}(2008)}]{antypov_computer_2008}%
  \BibitemOpen
  \bibfield  {author} {\bibinfo {author} {\bibfnamefont {D.}~\bibnamefont {Antypov}}\ and\ \bibinfo {author} {\bibfnamefont {J.~A.}\ \bibnamefont {Elliott}},\ }\bibfield  {title} {\enquote {\bibinfo {title} {Computer simulation study of a single polymer chain in an attractive solvent},}\ }\href {\doibase 10.1063/1.2991178} {\bibfield  {journal} {\bibinfo  {journal} {The Journal of Chemical Physics}\ }\textbf {\bibinfo {volume} {129}},\ \bibinfo {pages} {174901} (\bibinfo {year} {2008})}\BibitemShut {NoStop}%
\bibitem [{\citenamefont {Parsons}\ and\ \citenamefont {Williams}(2006)}]{parsons_off-lattice_2006}%
  \BibitemOpen
  \bibfield  {author} {\bibinfo {author} {\bibfnamefont {D.~F.}\ \bibnamefont {Parsons}}\ and\ \bibinfo {author} {\bibfnamefont {D.~R.~M.}\ \bibnamefont {Williams}},\ }\bibfield  {title} {\enquote {\bibinfo {title} {An off-lattice {Wang}-{Landau} study of the coil-globule and melting transitions of a flexible homopolymer},}\ }\href {\doibase 10.1063/1.2209684} {\bibfield  {journal} {\bibinfo  {journal} {The Journal of Chemical Physics}\ }\textbf {\bibinfo {volume} {124}},\ \bibinfo {pages} {221103} (\bibinfo {year} {2006})}\BibitemShut {NoStop}%
\bibitem [{\citenamefont {Boothroyd}\ \emph {et~al.}(2018)\citenamefont {Boothroyd}, \citenamefont {Kerridge}, \citenamefont {Broo}, \citenamefont {Buttar},\ and\ \citenamefont {Anwar}}]{boothroyd_solubility_2018}%
  \BibitemOpen
  \bibfield  {author} {\bibinfo {author} {\bibfnamefont {S.}~\bibnamefont {Boothroyd}}, \bibinfo {author} {\bibfnamefont {A.}~\bibnamefont {Kerridge}}, \bibinfo {author} {\bibfnamefont {A.}~\bibnamefont {Broo}}, \bibinfo {author} {\bibfnamefont {D.}~\bibnamefont {Buttar}}, \ and\ \bibinfo {author} {\bibfnamefont {J.}~\bibnamefont {Anwar}},\ }\bibfield  {title} {\enquote {\bibinfo {title} {Solubility prediction from first principles: a density of states approach},}\ }\href {\doibase 10.1039/C8CP01786G} {\bibfield  {journal} {\bibinfo  {journal} {Physical Chemistry Chemical Physics}\ }\textbf {\bibinfo {volume} {20}},\ \bibinfo {pages} {20981} (\bibinfo {year} {2018})}\BibitemShut {NoStop}%
\bibitem [{\citenamefont {Boothroyd}\ and\ \citenamefont {Anwar}(2019)}]{boothroyd_solubility_2019}%
  \BibitemOpen
  \bibfield  {author} {\bibinfo {author} {\bibfnamefont {S.}~\bibnamefont {Boothroyd}}\ and\ \bibinfo {author} {\bibfnamefont {J.}~\bibnamefont {Anwar}},\ }\bibfield  {title} {\enquote {\bibinfo {title} {Solubility prediction for a soluble organic molecule via chemical potentials from density of states},}\ }\href {\doibase 10.1063/1.5117281} {\bibfield  {journal} {\bibinfo  {journal} {The Journal of Chemical Physics}\ }\textbf {\bibinfo {volume} {151}},\ \bibinfo {pages} {184113} (\bibinfo {year} {2019})}\BibitemShut {NoStop}%
\bibitem [{\citenamefont {Bin-Omran}(2017)}]{bin-omran_influence_2017}%
  \BibitemOpen
  \bibfield  {author} {\bibinfo {author} {\bibfnamefont {S.}~\bibnamefont {Bin-Omran}},\ }\bibfield  {title} {\enquote {\bibinfo {title} {The influence of mechanical and electrical boundary conditions on electrocaloric response in ({Ba}\textsubscript{0.5}{Sr}\textsubscript{0.5}){TiO\textsubscript{3}} thin films},}\ }\href {\doibase 10.1016/j.materresbull.2017.07.049} {\bibfield  {journal} {\bibinfo  {journal} {Materials Research Bulletin}\ }\textbf {\bibinfo {volume} {95}},\ \bibinfo {pages} {334} (\bibinfo {year} {2017})}\BibitemShut {NoStop}%
\bibitem [{\citenamefont {Nguyen}\ and\ \citenamefont {Thanh~Ngo}(2017)}]{nguyen_study_2017}%
  \BibitemOpen
  \bibfield  {author} {\bibinfo {author} {\bibfnamefont {T.-L.~H.}\ \bibnamefont {Nguyen}}\ and\ \bibinfo {author} {\bibfnamefont {V.}~\bibnamefont {Thanh~Ngo}},\ }\bibfield  {title} {\enquote {\bibinfo {title} {Study on the critical properties of thin magnetic films using the clock model},}\ }\href {\doibase 10.1088/2043-6254/aa5981} {\bibfield  {journal} {\bibinfo  {journal} {Advances in Natural Sciences: Nanoscience and Nanotechnology}\ }\textbf {\bibinfo {volume} {8}},\ \bibinfo {pages} {015013} (\bibinfo {year} {2017})}\BibitemShut {NoStop}%
\bibitem [{\citenamefont {Miyashita}\ \emph {et~al.}(2021)\citenamefont {Miyashita}, \citenamefont {Nishino}, \citenamefont {Toga}, \citenamefont {Hinokihara}, \citenamefont {Uysal}, \citenamefont {Miyake}, \citenamefont {Akai}, \citenamefont {Hirosawa},\ and\ \citenamefont {Sakuma}}]{miyashita_atomistic_2021}%
  \BibitemOpen
  \bibfield  {author} {\bibinfo {author} {\bibfnamefont {S.}~\bibnamefont {Miyashita}}, \bibinfo {author} {\bibfnamefont {M.}~\bibnamefont {Nishino}}, \bibinfo {author} {\bibfnamefont {Y.}~\bibnamefont {Toga}}, \bibinfo {author} {\bibfnamefont {T.}~\bibnamefont {Hinokihara}}, \bibinfo {author} {\bibfnamefont {I.~E.}\ \bibnamefont {Uysal}}, \bibinfo {author} {\bibfnamefont {T.}~\bibnamefont {Miyake}}, \bibinfo {author} {\bibfnamefont {H.}~\bibnamefont {Akai}}, \bibinfo {author} {\bibfnamefont {S.}~\bibnamefont {Hirosawa}}, \ and\ \bibinfo {author} {\bibfnamefont {A.}~\bibnamefont {Sakuma}},\ }\bibfield  {title} {\enquote {\bibinfo {title} {Atomistic theory of thermally activated magnetization processes in {Nd\textsubscript{2}Fe\textsubscript{14}B} permanent magnet},}\ }\href {\doibase 10.1080/14686996.2021.1942197} {\bibfield  {journal} {\bibinfo  {journal} {Science and Technology of Advanced Materials}\ }\textbf {\bibinfo {volume} {22}},\ \bibinfo {pages} {658} (\bibinfo {year} {2021})}\BibitemShut {NoStop}%
\bibitem [{\citenamefont {Bin-Omran}\ \emph {et~al.}(2016)\citenamefont {Bin-Omran}, \citenamefont {Kornev},\ and\ \citenamefont {Bellaiche}}]{bin-omran_wang-landau_2016}%
  \BibitemOpen
  \bibfield  {author} {\bibinfo {author} {\bibfnamefont {S.}~\bibnamefont {Bin-Omran}}, \bibinfo {author} {\bibfnamefont {I.~A.}\ \bibnamefont {Kornev}}, \ and\ \bibinfo {author} {\bibfnamefont {L.}~\bibnamefont {Bellaiche}},\ }\bibfield  {title} {\enquote {\bibinfo {title} {Wang-{Landau} {Monte} {Carlo} formalism applied to ferroelectrics},}\ }\href {\doibase 10.1103/PhysRevB.93.014104} {\bibfield  {journal} {\bibinfo  {journal} {Physical Review B}\ }\textbf {\bibinfo {volume} {93}},\ \bibinfo {pages} {014104} (\bibinfo {year} {2016})}\BibitemShut {NoStop}%
\bibitem [{\citenamefont {Ngo}\ and\ \citenamefont {Diep}(2008)}]{ngo_phase_2008}%
  \BibitemOpen
  \bibfield  {author} {\bibinfo {author} {\bibfnamefont {V.~T.}\ \bibnamefont {Ngo}}\ and\ \bibinfo {author} {\bibfnamefont {H.~T.}\ \bibnamefont {Diep}},\ }\bibfield  {title} {\enquote {\bibinfo {title} {Phase transition in {Heisenberg} stacked triangular antiferromagnets: {End} of a controversy},}\ }\href {\doibase 10.1103/PhysRevE.78.031119} {\bibfield  {journal} {\bibinfo  {journal} {Physical Review E}\ }\textbf {\bibinfo {volume} {78}},\ \bibinfo {pages} {031119} (\bibinfo {year} {2008})}\BibitemShut {NoStop}%
\bibitem [{\citenamefont {Bogoslovskiy}\ \emph {et~al.}(2024)\citenamefont {Bogoslovskiy}, \citenamefont {Petrov},\ and\ \citenamefont {Averkiev}}]{bogoslovskiy_phase_2024}%
  \BibitemOpen
  \bibfield  {author} {\bibinfo {author} {\bibfnamefont {N.~A.}\ \bibnamefont {Bogoslovskiy}}, \bibinfo {author} {\bibfnamefont {P.~V.}\ \bibnamefont {Petrov}}, \ and\ \bibinfo {author} {\bibfnamefont {N.~S.}\ \bibnamefont {Averkiev}},\ }\bibfield  {title} {\enquote {\bibinfo {title} {Phase diagram of a ferromagnetic semiconductor: {Origin} of superparamagnetism},}\ }\href {\doibase 10.1103/PhysRevB.109.024436} {\bibfield  {journal} {\bibinfo  {journal} {Physical Review B}\ }\textbf {\bibinfo {volume} {109}},\ \bibinfo {pages} {024436} (\bibinfo {year} {2024})}\BibitemShut {NoStop}%
\bibitem [{\citenamefont {Allen}\ and\ \citenamefont {Swetnam}(2012)}]{allen_wang-landau_2012}%
  \BibitemOpen
  \bibfield  {author} {\bibinfo {author} {\bibfnamefont {M.~P.}\ \bibnamefont {Allen}}\ and\ \bibinfo {author} {\bibfnamefont {A.~D.}\ \bibnamefont {Swetnam}},\ }\bibfield  {title} {\enquote {\bibinfo {title} {Wang-{Landau} {Simulations} of {Adsorbed} and {Confined} {Lattice} {Polymers}},}\ }\href {\doibase 10.1016/j.phpro.2012.05.002} {\bibfield  {journal} {\bibinfo  {journal} {Physics Procedia}\ }\bibinfo {series} {Proceedings of the 25th {Workshop} on {Computer} {Simulation} {Studies} in {Condensed} {Matter} {Physics}},\ \textbf {\bibinfo {volume} {34}},\ \bibinfo {pages} {6} (\bibinfo {year} {2012})}\BibitemShut {NoStop}%
\bibitem [{\citenamefont {Lazo}\ and\ \citenamefont {Keil}(2009)}]{lazo_phase_2009}%
  \BibitemOpen
  \bibfield  {author} {\bibinfo {author} {\bibfnamefont {C.}~\bibnamefont {Lazo}}\ and\ \bibinfo {author} {\bibfnamefont {F.~J.}\ \bibnamefont {Keil}},\ }\bibfield  {title} {\enquote {\bibinfo {title} {Phase diagram of oxygen adsorbed on {Ni}(111) and thermodynamic properties from first-principles},}\ }\href {\doibase 10.1103/PhysRevB.79.245418} {\bibfield  {journal} {\bibinfo  {journal} {Physical Review B}\ }\textbf {\bibinfo {volume} {79}},\ \bibinfo {pages} {245418} (\bibinfo {year} {2009})}\BibitemShut {NoStop}%
\bibitem [{\citenamefont {Swetnam}\ and\ \citenamefont {Allen}(2009)}]{swetnam_improved_2009}%
  \BibitemOpen
  \bibfield  {author} {\bibinfo {author} {\bibfnamefont {A.~D.}\ \bibnamefont {Swetnam}}\ and\ \bibinfo {author} {\bibfnamefont {M.~P.}\ \bibnamefont {Allen}},\ }\bibfield  {title} {\enquote {\bibinfo {title} {Improved simulations of lattice peptide adsorption},}\ }\href {\doibase 10.1039/B818067A} {\bibfield  {journal} {\bibinfo  {journal} {Physical Chemistry Chemical Physics}\ }\textbf {\bibinfo {volume} {11}},\ \bibinfo {pages} {2046} (\bibinfo {year} {2009})}\BibitemShut {NoStop}%
\bibitem [{\citenamefont {Kim}\ \emph {et~al.}(2006)\citenamefont {Kim}, \citenamefont {Straub},\ and\ \citenamefont {Keyes}}]{kim_statistical-temperature_2006}%
  \BibitemOpen
  \bibfield  {author} {\bibinfo {author} {\bibfnamefont {J.}~\bibnamefont {Kim}}, \bibinfo {author} {\bibfnamefont {J.~E.}\ \bibnamefont {Straub}}, \ and\ \bibinfo {author} {\bibfnamefont {T.}~\bibnamefont {Keyes}},\ }\bibfield  {title} {\enquote {\bibinfo {title} {Statistical-{Temperature} {Monte} {Carlo} and {Molecular} {Dynamics} {Algorithms}},}\ }\href {\doibase 10.1103/PhysRevLett.97.050601} {\bibfield  {journal} {\bibinfo  {journal} {Physical Review Letters}\ }\textbf {\bibinfo {volume} {97}},\ \bibinfo {pages} {050601} (\bibinfo {year} {2006})}\BibitemShut {NoStop}%
\bibitem [{\citenamefont {Vogel}\ \emph {et~al.}(2013)\citenamefont {Vogel}, \citenamefont {Li}, \citenamefont {W\"ust},\ and\ \citenamefont {Landau}}]{wl_mpi_2013}%
  \BibitemOpen
  \bibfield  {author} {\bibinfo {author} {\bibfnamefont {T.}~\bibnamefont {Vogel}}, \bibinfo {author} {\bibfnamefont {Y.~W.}\ \bibnamefont {Li}}, \bibinfo {author} {\bibfnamefont {T.}~\bibnamefont {W\"ust}}, \ and\ \bibinfo {author} {\bibfnamefont {D.~P.}\ \bibnamefont {Landau}},\ }\bibfield  {title} {\enquote {\bibinfo {title} {{Generic}, {Hierarchical} {Framework} for {Massively} {Parallel} {Wang}-{Landau} {Sampling}},}\ }\href {\doibase 10.1103/PhysRevLett.110.210603} {\bibfield  {journal} {\bibinfo  {journal} {Physical Review Letters}\ }\textbf {\bibinfo {volume} {110}},\ \bibinfo {pages} {210603} (\bibinfo {year} {2013})}\BibitemShut {NoStop}%
\bibitem [{\citenamefont {Zierenberg}\ \emph {et~al.}(2013)\citenamefont {Zierenberg}, \citenamefont {Marenz},\ and\ \citenamefont {Janke}}]{zierenberg2013}%
  \BibitemOpen
  \bibfield  {author} {\bibinfo {author} {\bibfnamefont {J.}~\bibnamefont {Zierenberg}}, \bibinfo {author} {\bibfnamefont {M.}~\bibnamefont {Marenz}}, \ and\ \bibinfo {author} {\bibfnamefont {W.}~\bibnamefont {Janke}},\ }\bibfield  {title} {\enquote {\bibinfo {title} {Scaling properties of a parallel implementation of the multicanonical algorithm},}\ }\href {\doibase https://doi.org/10.1016/j.cpc.2012.12.006} {\bibfield  {journal} {\bibinfo  {journal} {Computer Physics Communications}\ }\textbf {\bibinfo {volume} {184}},\ \bibinfo {pages} {1155} (\bibinfo {year} {2013})}\BibitemShut {NoStop}%
\bibitem [{\citenamefont {Gross}\ \emph {et~al.}(2018)\citenamefont {Gross}, \citenamefont {Zierenberg}, \citenamefont {Weigel},\ and\ \citenamefont {Janke}}]{gross2018}%
  \BibitemOpen
  \bibfield  {author} {\bibinfo {author} {\bibfnamefont {J.}~\bibnamefont {Gross}}, \bibinfo {author} {\bibfnamefont {J.}~\bibnamefont {Zierenberg}}, \bibinfo {author} {\bibfnamefont {M.}~\bibnamefont {Weigel}}, \ and\ \bibinfo {author} {\bibfnamefont {W.}~\bibnamefont {Janke}},\ }\bibfield  {title} {\enquote {\bibinfo {title} {Massively parallel multicanonical simulations},}\ }\href {\doibase https://doi.org/10.1016/j.cpc.2017.10.018} {\bibfield  {journal} {\bibinfo  {journal} {Computer Physics Communications}\ }\textbf {\bibinfo {volume} {224}},\ \bibinfo {pages} {387} (\bibinfo {year} {2018})}\BibitemShut {NoStop}%
\bibitem [{\citenamefont {Zhou}\ and\ \citenamefont {Bhatt}(2005)}]{zhou_understanding_2005}%
  \BibitemOpen
  \bibfield  {author} {\bibinfo {author} {\bibfnamefont {C.}~\bibnamefont {Zhou}}\ and\ \bibinfo {author} {\bibfnamefont {R.~N.}\ \bibnamefont {Bhatt}},\ }\bibfield  {title} {\enquote {\bibinfo {title} {Understanding and improving the {Wang}-{Landau} algorithm},}\ }\href {\doibase 10.1103/PhysRevE.72.025701} {\bibfield  {journal} {\bibinfo  {journal} {Physical Review E}\ }\textbf {\bibinfo {volume} {72}},\ \bibinfo {pages} {025701} (\bibinfo {year} {2005})}\BibitemShut {NoStop}%
\bibitem [{\citenamefont {Belardinelli}\ and\ \citenamefont {Pereyra}(2007)}]{belardinelli_fast_2007}%
  \BibitemOpen
  \bibfield  {author} {\bibinfo {author} {\bibfnamefont {R.~E.}\ \bibnamefont {Belardinelli}}\ and\ \bibinfo {author} {\bibfnamefont {V.~D.}\ \bibnamefont {Pereyra}},\ }\bibfield  {title} {\enquote {\bibinfo {title} {Fast algorithm to calculate density of states},}\ }\href {\doibase 10.1103/PhysRevE.75.046701} {\bibfield  {journal} {\bibinfo  {journal} {Physical Review E}\ }\textbf {\bibinfo {volume} {75}},\ \bibinfo {pages} {046701} (\bibinfo {year} {2007})}\BibitemShut {NoStop}%
\bibitem [{\citenamefont {Zhou}\ \emph {et~al.}(2006)\citenamefont {Zhou}, \citenamefont {Schulthess}, \citenamefont {Torbrügge},\ and\ \citenamefont {Landau}}]{zhou_wang-landau_2006}%
  \BibitemOpen
  \bibfield  {author} {\bibinfo {author} {\bibfnamefont {C.}~\bibnamefont {Zhou}}, \bibinfo {author} {\bibfnamefont {T.~C.}\ \bibnamefont {Schulthess}}, \bibinfo {author} {\bibfnamefont {S.}~\bibnamefont {Torbrügge}}, \ and\ \bibinfo {author} {\bibfnamefont {D.~P.}\ \bibnamefont {Landau}},\ }\bibfield  {title} {\enquote {\bibinfo {title} {Wang-{Landau} {Algorithm} for {Continuous} {Models} and {Joint} {Density} of {States}},}\ }\href {\doibase 10.1103/PhysRevLett.96.120201} {\bibfield  {journal} {\bibinfo  {journal} {Physical Review Letters}\ }\textbf {\bibinfo {volume} {96}},\ \bibinfo {pages} {120201} (\bibinfo {year} {2006})}\BibitemShut {NoStop}%
\bibitem [{\citenamefont {Swetnam}\ and\ \citenamefont {Allen}(2011)}]{swetnam_improving_2011}%
  \BibitemOpen
  \bibfield  {author} {\bibinfo {author} {\bibfnamefont {A.~D.}\ \bibnamefont {Swetnam}}\ and\ \bibinfo {author} {\bibfnamefont {M.~P.}\ \bibnamefont {Allen}},\ }\bibfield  {title} {\enquote {\bibinfo {title} {Improving the {Wang}–{Landau} algorithm for polymers and proteins},}\ }\href {\doibase 10.1002/jcc.21660} {\bibfield  {journal} {\bibinfo  {journal} {Journal of Computational Chemistry}\ }\textbf {\bibinfo {volume} {32}},\ \bibinfo {pages} {816} (\bibinfo {year} {2011})}\BibitemShut {NoStop}%
\bibitem [{\citenamefont {Cunha-Netto}\ and\ \citenamefont {Dickman}(2011)}]{cunha-netto_critical_2011}%
  \BibitemOpen
  \bibfield  {author} {\bibinfo {author} {\bibfnamefont {A.~G.}\ \bibnamefont {Cunha-Netto}}\ and\ \bibinfo {author} {\bibfnamefont {R.}~\bibnamefont {Dickman}},\ }\bibfield  {title} {\enquote {\bibinfo {title} {Critical behavior of hard-core lattice gases: {Wang}–{Landau} sampling with adaptive windows},}\ }\href {\doibase 10.1016/j.cpc.2010.12.013} {\bibfield  {journal} {\bibinfo  {journal} {Computer Physics Communications}\ }\textbf {\bibinfo {volume} {182}},\ \bibinfo {pages} {719} (\bibinfo {year} {2011})}\BibitemShut {NoStop}%
\bibitem [{\citenamefont {Cunha-Netto}\ \emph {et~al.}(2008)\citenamefont {Cunha-Netto}, \citenamefont {Caparica}, \citenamefont {Tsai}, \citenamefont {Dickman},\ and\ \citenamefont {Landau}}]{cunha-netto_improving_2008}%
  \BibitemOpen
  \bibfield  {author} {\bibinfo {author} {\bibfnamefont {A.~G.}\ \bibnamefont {Cunha-Netto}}, \bibinfo {author} {\bibfnamefont {A.~A.}\ \bibnamefont {Caparica}}, \bibinfo {author} {\bibfnamefont {S.-H.}\ \bibnamefont {Tsai}}, \bibinfo {author} {\bibfnamefont {R.}~\bibnamefont {Dickman}}, \ and\ \bibinfo {author} {\bibfnamefont {D.~P.}\ \bibnamefont {Landau}},\ }\bibfield  {title} {\enquote {\bibinfo {title} {Improving {Wang}-{Landau} sampling with adaptive windows},}\ }\href {\doibase 10.1103/PhysRevE.78.055701} {\bibfield  {journal} {\bibinfo  {journal} {Physical Review E}\ }\textbf {\bibinfo {volume} {78}},\ \bibinfo {pages} {055701} (\bibinfo {year} {2008})}\BibitemShut {NoStop}%
\bibitem [{\citenamefont {Cunha-Netto}\ \emph {et~al.}(2009)\citenamefont {Cunha-Netto}, \citenamefont {Dickman},\ and\ \citenamefont {Caparica}}]{cunha-netto_two-dimensional_2009}%
  \BibitemOpen
  \bibfield  {author} {\bibinfo {author} {\bibfnamefont {A.~G.}\ \bibnamefont {Cunha-Netto}}, \bibinfo {author} {\bibfnamefont {R.}~\bibnamefont {Dickman}}, \ and\ \bibinfo {author} {\bibfnamefont {A.~A.}\ \bibnamefont {Caparica}},\ }\bibfield  {title} {\enquote {\bibinfo {title} {Two-dimensional lattice polymers: {Adaptive} windows simulations},}\ }\href {\doibase 10.1016/j.cpc.2008.12.015} {\bibfield  {journal} {\bibinfo  {journal} {Computer Physics Communications}\ }\bibinfo {series} {Special issue based on the {Conference} on {Computational} {Physics} 2008},\ \textbf {\bibinfo {volume} {180}},\ \bibinfo {pages} {583} (\bibinfo {year} {2009})}\BibitemShut {NoStop}%
\bibitem [{\citenamefont {Yin}\ and\ \citenamefont {Landau}(2012)}]{junqi2012parallel}%
  \BibitemOpen
  \bibfield  {author} {\bibinfo {author} {\bibfnamefont {J.}~\bibnamefont {Yin}}\ and\ \bibinfo {author} {\bibfnamefont {D.}~\bibnamefont {Landau}},\ }\bibfield  {title} {\enquote {\bibinfo {title} {Massively parallel {Wang}–{Landau} sampling on multiple {GPUs}},}\ }\href {\doibase https://doi.org/10.1016/j.cpc.2012.02.023} {\bibfield  {journal} {\bibinfo  {journal} {Computer Physics Communications}\ }\textbf {\bibinfo {volume} {183}},\ \bibinfo {pages} {1568} (\bibinfo {year} {2012})}\BibitemShut {NoStop}%
\bibitem [{\citenamefont {Vogel}\ \emph {et~al.}(2014)\citenamefont {Vogel}, \citenamefont {Li}, \citenamefont {W\"ust},\ and\ \citenamefont {Landau}}]{vogel2014replica}%
  \BibitemOpen
  \bibfield  {author} {\bibinfo {author} {\bibfnamefont {T.}~\bibnamefont {Vogel}}, \bibinfo {author} {\bibfnamefont {Y.~W.}\ \bibnamefont {Li}}, \bibinfo {author} {\bibfnamefont {T.}~\bibnamefont {W\"ust}}, \ and\ \bibinfo {author} {\bibfnamefont {D.~P.}\ \bibnamefont {Landau}},\ }\bibfield  {title} {\enquote {\bibinfo {title} {Scalable replica-exchange framework for {Wang}-{Landau} sampling},}\ }\href {\doibase 10.1103/PhysRevE.90.023302} {\bibfield  {journal} {\bibinfo  {journal} {Physical Review E}\ }\textbf {\bibinfo {volume} {90}},\ \bibinfo {pages} {023302} (\bibinfo {year} {2014})}\BibitemShut {NoStop}%
\bibitem [{\citenamefont {Pei}(2019)}]{pei2019fluctuation}%
  \BibitemOpen
  \bibfield  {author} {\bibinfo {author} {\bibfnamefont {Z.}~\bibnamefont {Pei}},\ }\bibfield  {title} {\enquote {\bibinfo {title} {Theory of the energy fluctuation of multicomponent alloys},}\ }\href {\doibase https://doi.org/10.1016/j.scriptamat.2018.12.004} {\bibfield  {journal} {\bibinfo  {journal} {Scripta Materialia}\ }\textbf {\bibinfo {volume} {162}},\ \bibinfo {pages} {503} (\bibinfo {year} {2019})}\BibitemShut {NoStop}%
\bibitem [{\citenamefont {Geyer}(1991)}]{geyer_markov_1991}%
  \BibitemOpen
  \bibfield  {author} {\bibinfo {author} {\bibfnamefont {C.~J.}\ \bibnamefont {Geyer}},\ }\href {https://hdl.handle.net/11299/58440} {\emph {\bibinfo {title} {Markov Chain Monte Carlo Maximum Likelihood}}},\ \bibinfo {type} {Tech. Rep.}\ \bibinfo {number} {Technical Report No. 568}\ (\bibinfo  {institution} {University of Minnesota, School of Statistics},\ \bibinfo {year} {1991})\BibitemShut {NoStop}%
\bibitem [{\citenamefont {Hukushima}\ and\ \citenamefont {Nemoto}(1996)}]{hukushima_exchange_1996}%
  \BibitemOpen
  \bibfield  {author} {\bibinfo {author} {\bibfnamefont {K.}~\bibnamefont {Hukushima}}\ and\ \bibinfo {author} {\bibfnamefont {K.}~\bibnamefont {Nemoto}},\ }\bibfield  {title} {\enquote {\bibinfo {title} {Exchange {Monte} {Carlo} {Method} and {Application} to {Spin} {Glass} {Simulations}},}\ }\href {\doibase 10.1143/JPSJ.65.1604} {\bibfield  {journal} {\bibinfo  {journal} {Journal of the Physical Society of Japan}\ }\textbf {\bibinfo {volume} {65}},\ \bibinfo {pages} {1604} (\bibinfo {year} {1996})}\BibitemShut {NoStop}%
\bibitem [{\citenamefont {Valentim}\ \emph {et~al.}(2015)\citenamefont {Valentim}, \citenamefont {Rocha}, \citenamefont {Tsai}, \citenamefont {Li}, \citenamefont {Eisenbach}, \citenamefont {Fiore},\ and\ \citenamefont {Landau}}]{valentim_exploring_2015}%
  \BibitemOpen
  \bibfield  {author} {\bibinfo {author} {\bibfnamefont {A.}~\bibnamefont {Valentim}}, \bibinfo {author} {\bibfnamefont {J.~C.~S.}\ \bibnamefont {Rocha}}, \bibinfo {author} {\bibfnamefont {S.-H.}\ \bibnamefont {Tsai}}, \bibinfo {author} {\bibfnamefont {Y.~W.}\ \bibnamefont {Li}}, \bibinfo {author} {\bibfnamefont {M.}~\bibnamefont {Eisenbach}}, \bibinfo {author} {\bibfnamefont {C.~E.}\ \bibnamefont {Fiore}}, \ and\ \bibinfo {author} {\bibfnamefont {D.~P.}\ \bibnamefont {Landau}},\ }\bibfield  {title} {\enquote {\bibinfo {title} {Exploring {Replica}-{Exchange} {Wang}-{Landau} sampling in higher-dimensional parameter space},}\ }\href {\doibase 10.1088/1742-6596/640/1/012006} {\bibfield  {journal} {\bibinfo  {journal} {Journal of Physics: Conference Series}\ }\textbf {\bibinfo {volume} {640}},\ \bibinfo {pages} {012006} (\bibinfo {year} {2015})}\BibitemShut {NoStop}%
\bibitem [{\citenamefont {Zhao}\ \emph {et~al.}(2014)\citenamefont {Zhao}, \citenamefont {Cheung}, \citenamefont {Li},\ and\ \citenamefont {Eisenbach}}]{zhao_performance_2014}%
  \BibitemOpen
  \bibfield  {author} {\bibinfo {author} {\bibfnamefont {Y.}~\bibnamefont {Zhao}}, \bibinfo {author} {\bibfnamefont {S.~W.}\ \bibnamefont {Cheung}}, \bibinfo {author} {\bibfnamefont {Y.~W.}\ \bibnamefont {Li}}, \ and\ \bibinfo {author} {\bibfnamefont {M.}~\bibnamefont {Eisenbach}},\ }\bibfield  {title} {\enquote {\bibinfo {title} {{Performance} of {Replica}-{exchange} {Wang}-{Landau} {Sampling} for the {2D} {Ising} {Model}: {A} {Brief} {Survey}},}\ }\href {\doibase 10.1016/j.phpro.2014.08.129} {\bibfield  {journal} {\bibinfo  {journal} {Physics Procedia}\ }\bibinfo {series} {Proceedings of the 27th Workshop on Computer Simulation Studies in Condensed Matter Physics ({CSP}2014)},\ \textbf {\bibinfo {volume} {57}},\ \bibinfo {pages} {43} (\bibinfo {year} {2014})}\BibitemShut {NoStop}%
\bibitem [{\citenamefont {K{\"a}stner}(2011)}]{kastner2011umbrella}%
  \BibitemOpen
  \bibfield  {author} {\bibinfo {author} {\bibfnamefont {J.}~\bibnamefont {K{\"a}stner}},\ }\bibfield  {title} {\enquote {\bibinfo {title} {Umbrella sampling},}\ }\href@noop {} {\bibfield  {journal} {\bibinfo  {journal} {Wiley Interdisciplinary Reviews: Computational Molecular Science}\ }\textbf {\bibinfo {volume} {1}},\ \bibinfo {pages} {932} (\bibinfo {year} {2011})}\BibitemShut {NoStop}%
\bibitem [{\citenamefont {Khan}\ \emph {et~al.}(2016)\citenamefont {Khan}, \citenamefont {Staunton},\ and\ \citenamefont {Stocks}}]{khan_statistical_2016}%
  \BibitemOpen
  \bibfield  {author} {\bibinfo {author} {\bibfnamefont {S.~N.}\ \bibnamefont {Khan}}, \bibinfo {author} {\bibfnamefont {J.~B.}\ \bibnamefont {Staunton}}, \ and\ \bibinfo {author} {\bibfnamefont {G.~M.}\ \bibnamefont {Stocks}},\ }\bibfield  {title} {\enquote {\bibinfo {title} {Statistical physics of multicomponent alloys using {KKR}-{CPA}},}\ }\href {\doibase 10.1103/PhysRevB.93.054206} {\bibfield  {journal} {\bibinfo  {journal} {Physical Review B}\ }\textbf {\bibinfo {volume} {93}},\ \bibinfo {pages} {054206} (\bibinfo {year} {2016})}\BibitemShut {NoStop}%
\bibitem [{\citenamefont {Naguszewski}\ \emph {et~al.}(2025)\citenamefont {Naguszewski}, \citenamefont {Pártay}, \citenamefont {Quigley},\ and\ \citenamefont {Woodgate}}]{naguszewski_BraWl_2025}%
  \BibitemOpen
  \bibfield  {author} {\bibinfo {author} {\bibfnamefont {H.~J.}\ \bibnamefont {Naguszewski}}, \bibinfo {author} {\bibfnamefont {L.~B.}\ \bibnamefont {Pártay}}, \bibinfo {author} {\bibfnamefont {D.}~\bibnamefont {Quigley}}, \ and\ \bibinfo {author} {\bibfnamefont {C.~D.}\ \bibnamefont {Woodgate}},\ }\bibfield  {title} {\enquote {\bibinfo {title} {{BraWl}: {Simulating} the thermodynamics and phase stability of multicomponent alloys using conventional and enhanced sampling techniques},}\ }\href {\doibase 10.21105/joss.08346} {\bibfield  {journal} {\bibinfo  {journal} {Journal of Open Source Software}\ }\textbf {\bibinfo {volume} {10}},\ \bibinfo {pages} {8346} (\bibinfo {year} {2025})}\BibitemShut {NoStop}%
\bibitem [{\citenamefont {Khan}\ and\ \citenamefont {Eisenbach}(2016)}]{khan_density-functional_2016}%
  \BibitemOpen
  \bibfield  {author} {\bibinfo {author} {\bibfnamefont {S.~N.}\ \bibnamefont {Khan}}\ and\ \bibinfo {author} {\bibfnamefont {M.}~\bibnamefont {Eisenbach}},\ }\bibfield  {title} {\enquote {\bibinfo {title} {Density-functional {Monte}-{Carlo} simulation of {CuZn} order-disorder transition},}\ }\href {\doibase 10.1103/PhysRevB.93.024203} {\bibfield  {journal} {\bibinfo  {journal} {Physical Review B}\ }\textbf {\bibinfo {volume} {93}},\ \bibinfo {pages} {024203} (\bibinfo {year} {2016})}\BibitemShut {NoStop}%
\bibitem [{\citenamefont {Pei}\ \emph {et~al.}(2020)\citenamefont {Pei}, \citenamefont {Li}, \citenamefont {Gao},\ and\ \citenamefont {Stocks}}]{pei_statistics_2020}%
  \BibitemOpen
  \bibfield  {author} {\bibinfo {author} {\bibfnamefont {Z.}~\bibnamefont {Pei}}, \bibinfo {author} {\bibfnamefont {R.}~\bibnamefont {Li}}, \bibinfo {author} {\bibfnamefont {M.~C.}\ \bibnamefont {Gao}}, \ and\ \bibinfo {author} {\bibfnamefont {G.~M.}\ \bibnamefont {Stocks}},\ }\bibfield  {title} {\enquote {\bibinfo {title} {Statistics of the {NiCoCr} medium-entropy alloy: {Novel} aspects of an old puzzle},}\ }\href {\doibase 10.1038/s41524-020-00389-1} {\bibfield  {journal} {\bibinfo  {journal} {npj Computational Materials}\ }\textbf {\bibinfo {volume} {6}},\ \bibinfo {pages} {122} (\bibinfo {year} {2020})}\BibitemShut {NoStop}%
\bibitem [{\citenamefont {Takeuchi}\ \emph {et~al.}(2017)\citenamefont {Takeuchi}, \citenamefont {Tanaka},\ and\ \citenamefont {Yuge}}]{takeuchi_new_2017}%
  \BibitemOpen
  \bibfield  {author} {\bibinfo {author} {\bibfnamefont {K.}~\bibnamefont {Takeuchi}}, \bibinfo {author} {\bibfnamefont {R.}~\bibnamefont {Tanaka}}, \ and\ \bibinfo {author} {\bibfnamefont {K.}~\bibnamefont {Yuge}},\ }\bibfield  {title} {\enquote {\bibinfo {title} {New {Wang}-{Landau} approach to obtain phase diagrams for multicomponent alloys},}\ }\href {\doibase 10.1103/PhysRevB.96.144202} {\bibfield  {journal} {\bibinfo  {journal} {Physical Review B}\ }\textbf {\bibinfo {volume} {96}},\ \bibinfo {pages} {144202} (\bibinfo {year} {2017})}\BibitemShut {NoStop}%
\bibitem [{\citenamefont {Nanba}\ and\ \citenamefont {Koyama}(2023)}]{nanba2023implementation}%
  \BibitemOpen
  \bibfield  {author} {\bibinfo {author} {\bibfnamefont {Y.}~\bibnamefont {Nanba}}\ and\ \bibinfo {author} {\bibfnamefont {M.}~\bibnamefont {Koyama}},\ }\bibfield  {title} {\enquote {\bibinfo {title} {{Implementation} of {Wang}-{Landau} {Algorithm} for {Probing} {Thermodynamic} {Stable} {Configuration} of {Multi}-{Element} {Materials} and {Application} to {Multinary} {Alloy} {Nanoparticles}},}\ }\href@noop {} {\bibfield  {journal} {\bibinfo  {journal} {Journal of Computer Chemistry, Japan-International Edition}\ }\textbf {\bibinfo {volume} {9}},\ \bibinfo {pages} {2022} (\bibinfo {year} {2023})}\BibitemShut {NoStop}%
\bibitem [{\citenamefont {Yeh}\ \emph {et~al.}(2004)\citenamefont {Yeh}, \citenamefont {Chen}, \citenamefont {Lin}, \citenamefont {Gan}, \citenamefont {Chin}, \citenamefont {Shun}, \citenamefont {Tsau},\ and\ \citenamefont {Chang}}]{yeh_nanostructured_2004}%
  \BibitemOpen
  \bibfield  {author} {\bibinfo {author} {\bibfnamefont {J.-W.}\ \bibnamefont {Yeh}}, \bibinfo {author} {\bibfnamefont {S.-K.}\ \bibnamefont {Chen}}, \bibinfo {author} {\bibfnamefont {S.-J.}\ \bibnamefont {Lin}}, \bibinfo {author} {\bibfnamefont {J.-Y.}\ \bibnamefont {Gan}}, \bibinfo {author} {\bibfnamefont {T.-S.}\ \bibnamefont {Chin}}, \bibinfo {author} {\bibfnamefont {T.-T.}\ \bibnamefont {Shun}}, \bibinfo {author} {\bibfnamefont {C.-H.}\ \bibnamefont {Tsau}}, \ and\ \bibinfo {author} {\bibfnamefont {S.-Y.}\ \bibnamefont {Chang}},\ }\bibfield  {title} {\enquote {\bibinfo {title} {Nanostructured {High}-{Entropy} {Alloys} with {Multiple} {Principal} {Elements}: {Novel} {Alloy} {Design} {Concepts} and {Outcomes}},}\ }\href {\doibase 10.1002/adem.200300567} {\bibfield  {journal} {\bibinfo  {journal} {Advanced Engineering Materials}\ }\textbf {\bibinfo {volume} {6}},\ \bibinfo {pages} {299} (\bibinfo {year} {2004})}\BibitemShut {NoStop}%
\bibitem [{\citenamefont {Widom}(2018)}]{widom_modeling_2018}%
  \BibitemOpen
  \bibfield  {author} {\bibinfo {author} {\bibfnamefont {M.}~\bibnamefont {Widom}},\ }\bibfield  {title} {\enquote {\bibinfo {title} {Modeling the structure and thermodynamics of high-entropy alloys},}\ }\href {\doibase 10.1557/jmr.2018.222} {\bibfield  {journal} {\bibinfo  {journal} {Journal of Materials Research}\ }\textbf {\bibinfo {volume} {33}},\ \bibinfo {pages} {2881} (\bibinfo {year} {2018})}\BibitemShut {NoStop}%
\bibitem [{\citenamefont {Eisenbach}\ \emph {et~al.}(2019)\citenamefont {Eisenbach}, \citenamefont {Pei},\ and\ \citenamefont {Liu}}]{eisenbach_first-principles_2019}%
  \BibitemOpen
  \bibfield  {author} {\bibinfo {author} {\bibfnamefont {M.}~\bibnamefont {Eisenbach}}, \bibinfo {author} {\bibfnamefont {Z.}~\bibnamefont {Pei}}, \ and\ \bibinfo {author} {\bibfnamefont {X.}~\bibnamefont {Liu}},\ }\bibfield  {title} {\enquote {\bibinfo {title} {First-principles study of order-disorder transitions in multicomponent solid-solution alloys},}\ }\href {\doibase 10.1088/1361-648X/ab13d8} {\bibfield  {journal} {\bibinfo  {journal} {Journal of Physics: Condensed Matter}\ }\textbf {\bibinfo {volume} {31}},\ \bibinfo {pages} {273002} (\bibinfo {year} {2019})}\BibitemShut {NoStop}%
\bibitem [{\citenamefont {Ferrari}\ \emph {et~al.}(2020)\citenamefont {Ferrari}, \citenamefont {Dutta}, \citenamefont {Gubaev}, \citenamefont {Ikeda}, \citenamefont {Srinivasan}, \citenamefont {Grabowski},\ and\ \citenamefont {Körmann}}]{ferrari_frontiers_2020}%
  \BibitemOpen
  \bibfield  {author} {\bibinfo {author} {\bibfnamefont {A.}~\bibnamefont {Ferrari}}, \bibinfo {author} {\bibfnamefont {B.}~\bibnamefont {Dutta}}, \bibinfo {author} {\bibfnamefont {K.}~\bibnamefont {Gubaev}}, \bibinfo {author} {\bibfnamefont {Y.}~\bibnamefont {Ikeda}}, \bibinfo {author} {\bibfnamefont {P.}~\bibnamefont {Srinivasan}}, \bibinfo {author} {\bibfnamefont {B.}~\bibnamefont {Grabowski}}, \ and\ \bibinfo {author} {\bibfnamefont {F.}~\bibnamefont {Körmann}},\ }\bibfield  {title} {\enquote {\bibinfo {title} {Frontiers in atomistic simulations of high entropy alloys},}\ }\href {\doibase 10.1063/5.0025310} {\bibfield  {journal} {\bibinfo  {journal} {Journal of Applied Physics}\ }\textbf {\bibinfo {volume} {128}},\ \bibinfo {pages} {150901} (\bibinfo {year} {2020})}\BibitemShut {NoStop}%
\bibitem [{\citenamefont {Ferrari}\ \emph {et~al.}(2023)\citenamefont {Ferrari}, \citenamefont {Körmann}, \citenamefont {Asta},\ and\ \citenamefont {Neugebauer}}]{ferrari_simulating_2023}%
  \BibitemOpen
  \bibfield  {author} {\bibinfo {author} {\bibfnamefont {A.}~\bibnamefont {Ferrari}}, \bibinfo {author} {\bibfnamefont {F.}~\bibnamefont {Körmann}}, \bibinfo {author} {\bibfnamefont {M.}~\bibnamefont {Asta}}, \ and\ \bibinfo {author} {\bibfnamefont {J.}~\bibnamefont {Neugebauer}},\ }\bibfield  {title} {\enquote {\bibinfo {title} {Simulating short-range order in compositionally complex materials},}\ }\href {\doibase 10.1038/s43588-023-00407-4} {\bibfield  {journal} {\bibinfo  {journal} {Nature Computational Science}\ }\textbf {\bibinfo {volume} {3}},\ \bibinfo {pages} {221} (\bibinfo {year} {2023})}\BibitemShut {NoStop}%
\bibitem [{\citenamefont {Zhang}\ \emph {et~al.}(2025)\citenamefont {Zhang}, \citenamefont {Sorkin}, \citenamefont {Aitken}, \citenamefont {Politano}, \citenamefont {Behler}, \citenamefont {P~Thompson}, \citenamefont {Ko}, \citenamefont {Ong}, \citenamefont {Chalykh}, \citenamefont {Korogod}, \citenamefont {Podryabinkin}, \citenamefont {Shapeev}, \citenamefont {Li}, \citenamefont {Mishin}, \citenamefont {Pei}, \citenamefont {Liu}, \citenamefont {Kim}, \citenamefont {Park}, \citenamefont {Hwang}, \citenamefont {Han}, \citenamefont {Sheriff}, \citenamefont {Cao},\ and\ \citenamefont {Freitas}}]{zhang_roadmap_2025}%
  \BibitemOpen
  \bibfield  {author} {\bibinfo {author} {\bibfnamefont {Y.-W.}\ \bibnamefont {Zhang}}, \bibinfo {author} {\bibfnamefont {V.}~\bibnamefont {Sorkin}}, \bibinfo {author} {\bibfnamefont {Z.~H.}\ \bibnamefont {Aitken}}, \bibinfo {author} {\bibfnamefont {A.}~\bibnamefont {Politano}}, \bibinfo {author} {\bibfnamefont {J.}~\bibnamefont {Behler}}, \bibinfo {author} {\bibfnamefont {A.}~\bibnamefont {P~Thompson}}, \bibinfo {author} {\bibfnamefont {T.~W.}\ \bibnamefont {Ko}}, \bibinfo {author} {\bibfnamefont {S.~P.}\ \bibnamefont {Ong}}, \bibinfo {author} {\bibfnamefont {O.}~\bibnamefont {Chalykh}}, \bibinfo {author} {\bibfnamefont {D.}~\bibnamefont {Korogod}}, \bibinfo {author} {\bibfnamefont {E.}~\bibnamefont {Podryabinkin}}, \bibinfo {author} {\bibfnamefont {A.}~\bibnamefont {Shapeev}}, \bibinfo {author} {\bibfnamefont {J.}~\bibnamefont {Li}}, \bibinfo {author} {\bibfnamefont {Y.}~\bibnamefont {Mishin}}, \bibinfo {author} {\bibfnamefont {Z.}~\bibnamefont {Pei}}, \bibinfo {author} {\bibfnamefont {X.}~\bibnamefont
  {Liu}}, \bibinfo {author} {\bibfnamefont {J.}~\bibnamefont {Kim}}, \bibinfo {author} {\bibfnamefont {Y.}~\bibnamefont {Park}}, \bibinfo {author} {\bibfnamefont {S.}~\bibnamefont {Hwang}}, \bibinfo {author} {\bibfnamefont {S.}~\bibnamefont {Han}}, \bibinfo {author} {\bibfnamefont {K.}~\bibnamefont {Sheriff}}, \bibinfo {author} {\bibfnamefont {Y.}~\bibnamefont {Cao}}, \ and\ \bibinfo {author} {\bibfnamefont {R.}~\bibnamefont {Freitas}},\ }\bibfield  {title} {\enquote {\bibinfo {title} {Roadmap for the development of machine learning-based interatomic potentials},}\ }\href {\doibase 10.1088/1361-651X/ad9d63} {\bibfield  {journal} {\bibinfo  {journal} {Modelling and Simulation in Materials Science and Engineering}\ }\textbf {\bibinfo {volume} {33}},\ \bibinfo {pages} {023301} (\bibinfo {year} {2025})}\BibitemShut {NoStop}%
\bibitem [{\citenamefont {Miracle}\ \emph {et~al.}(2020)\citenamefont {Miracle}, \citenamefont {Tsai}, \citenamefont {Senkov}, \citenamefont {Soni},\ and\ \citenamefont {Banerjee}}]{miracle_refractory_2020}%
  \BibitemOpen
  \bibfield  {author} {\bibinfo {author} {\bibfnamefont {D.~B.}\ \bibnamefont {Miracle}}, \bibinfo {author} {\bibfnamefont {M.-H.}\ \bibnamefont {Tsai}}, \bibinfo {author} {\bibfnamefont {O.~N.}\ \bibnamefont {Senkov}}, \bibinfo {author} {\bibfnamefont {V.}~\bibnamefont {Soni}}, \ and\ \bibinfo {author} {\bibfnamefont {R.}~\bibnamefont {Banerjee}},\ }\bibfield  {title} {\enquote {\bibinfo {title} {Refractory high entropy superalloys ({RSAs})},}\ }\href {\doibase 10.1016/j.scriptamat.2020.06.048} {\bibfield  {journal} {\bibinfo  {journal} {Scripta Materialia}\ }\textbf {\bibinfo {volume} {187}},\ \bibinfo {pages} {445} (\bibinfo {year} {2020})}\BibitemShut {NoStop}%
\bibitem [{\citenamefont {Chen}\ \emph {et~al.}(2019)\citenamefont {Chen}, \citenamefont {Kauffmann}, \citenamefont {Seils}, \citenamefont {Boll}, \citenamefont {Liebscher}, \citenamefont {Harding}, \citenamefont {Kumar}, \citenamefont {Szabó}, \citenamefont {Schlabach}, \citenamefont {Kauffmann-Weiss}, \citenamefont {Müller}, \citenamefont {Gorr}, \citenamefont {Christ},\ and\ \citenamefont {Heilmaier}}]{chen_crystallographic_2019}%
  \BibitemOpen
  \bibfield  {author} {\bibinfo {author} {\bibfnamefont {H.}~\bibnamefont {Chen}}, \bibinfo {author} {\bibfnamefont {A.}~\bibnamefont {Kauffmann}}, \bibinfo {author} {\bibfnamefont {S.}~\bibnamefont {Seils}}, \bibinfo {author} {\bibfnamefont {T.}~\bibnamefont {Boll}}, \bibinfo {author} {\bibfnamefont {C.}~\bibnamefont {Liebscher}}, \bibinfo {author} {\bibfnamefont {I.}~\bibnamefont {Harding}}, \bibinfo {author} {\bibfnamefont {K.}~\bibnamefont {Kumar}}, \bibinfo {author} {\bibfnamefont {D.}~\bibnamefont {Szabó}}, \bibinfo {author} {\bibfnamefont {S.}~\bibnamefont {Schlabach}}, \bibinfo {author} {\bibfnamefont {S.}~\bibnamefont {Kauffmann-Weiss}}, \bibinfo {author} {\bibfnamefont {F.}~\bibnamefont {Müller}}, \bibinfo {author} {\bibfnamefont {B.}~\bibnamefont {Gorr}}, \bibinfo {author} {\bibfnamefont {H.-J.}\ \bibnamefont {Christ}}, \ and\ \bibinfo {author} {\bibfnamefont {M.}~\bibnamefont {Heilmaier}},\ }\bibfield  {title} {\enquote {\bibinfo {title} {Crystallographic ordering in a series of {Al}-containing
  refractory high entropy alloys {Ta}–{Nb}–{Mo}–{Cr}–{Ti}–{Al}},}\ }\href {\doibase 10.1016/j.actamat.2019.07.001} {\bibfield  {journal} {\bibinfo  {journal} {Acta Materialia}\ }\textbf {\bibinfo {volume} {176}},\ \bibinfo {pages} {123} (\bibinfo {year} {2019})}\BibitemShut {NoStop}%
\bibitem [{\citenamefont {Woodgate}\ \emph {et~al.}(2025)\citenamefont {Woodgate}, \citenamefont {Naguszewski}, \citenamefont {Redka}, \citenamefont {Minár}, \citenamefont {Quigley},\ and\ \citenamefont {Staunton}}]{woodgate_emergent_2025}%
  \BibitemOpen
  \bibfield  {author} {\bibinfo {author} {\bibfnamefont {C.~D.}\ \bibnamefont {Woodgate}}, \bibinfo {author} {\bibfnamefont {H.~J.}\ \bibnamefont {Naguszewski}}, \bibinfo {author} {\bibfnamefont {D.}~\bibnamefont {Redka}}, \bibinfo {author} {\bibfnamefont {J.}~\bibnamefont {Minár}}, \bibinfo {author} {\bibfnamefont {D.}~\bibnamefont {Quigley}}, \ and\ \bibinfo {author} {\bibfnamefont {J.~B.}\ \bibnamefont {Staunton}},\ }\bibfield  {title} {\enquote {\bibinfo {title} {Emergent {B2} chemical orderings in the {AlTiVNb} and {AlTiCrMo} refractory high-entropy superalloys studied via first-principles theory and atomistic modelling},}\ }\href {\doibase 10.1088/2515-7639/adf468} {\bibfield  {journal} {\bibinfo  {journal} {Journal of Physics: Materials}\ }\textbf {\bibinfo {volume} {8}},\ \bibinfo {pages} {045002} (\bibinfo {year} {2025})}\BibitemShut {NoStop}%
\bibitem [{\citenamefont {Bragg}\ and\ \citenamefont {Williams}(1934)}]{bragg_effect_1934}%
  \BibitemOpen
  \bibfield  {author} {\bibinfo {author} {\bibfnamefont {W.~L.}\ \bibnamefont {Bragg}}\ and\ \bibinfo {author} {\bibfnamefont {E.~J.}\ \bibnamefont {Williams}},\ }\bibfield  {title} {\enquote {\bibinfo {title} {The effect of thermal agitation on atomic arrangement in alloys},}\ }\href {\doibase 10.1098/rspa.1934.0132} {\bibfield  {journal} {\bibinfo  {journal} {Proceedings of the Royal Society of London. Series A, Containing Papers of a Mathematical and Physical Character}\ }\textbf {\bibinfo {volume} {145}},\ \bibinfo {pages} {699} (\bibinfo {year} {1934})}\BibitemShut {NoStop}%
\bibitem [{\citenamefont {Bragg}\ and\ \citenamefont {Williams}(1935)}]{bragg_effect_1935}%
  \BibitemOpen
  \bibfield  {author} {\bibinfo {author} {\bibfnamefont {W.~L.}\ \bibnamefont {Bragg}}\ and\ \bibinfo {author} {\bibfnamefont {E.~J.}\ \bibnamefont {Williams}},\ }\bibfield  {title} {\enquote {\bibinfo {title} {The effect of thermal agitaion on atomic arrangement in alloys-{II}},}\ }\href {\doibase 10.1098/rspa.1935.0165} {\bibfield  {journal} {\bibinfo  {journal} {Proceedings of the Royal Society of London. Series A - Mathematical and Physical Sciences}\ }\textbf {\bibinfo {volume} {151}},\ \bibinfo {pages} {540} (\bibinfo {year} {1935})}\BibitemShut {NoStop}%
\bibitem [{\citenamefont {Brush}(1967)}]{brush_history_1967}%
  \BibitemOpen
  \bibfield  {author} {\bibinfo {author} {\bibfnamefont {S.~G.}\ \bibnamefont {Brush}},\ }\bibfield  {title} {\enquote {\bibinfo {title} {History of the {Lenz}-{Ising} {Model}},}\ }\href {\doibase 10.1103/RevModPhys.39.883} {\bibfield  {journal} {\bibinfo  {journal} {Reviews of Modern Physics}\ }\textbf {\bibinfo {volume} {39}},\ \bibinfo {pages} {883} (\bibinfo {year} {1967})}\BibitemShut {NoStop}%
\bibitem [{\citenamefont {Woodgate}\ and\ \citenamefont {Staunton}(2022)}]{woodgate_compositional_2022}%
  \BibitemOpen
  \bibfield  {author} {\bibinfo {author} {\bibfnamefont {C.~D.}\ \bibnamefont {Woodgate}}\ and\ \bibinfo {author} {\bibfnamefont {J.~B.}\ \bibnamefont {Staunton}},\ }\bibfield  {title} {\enquote {\bibinfo {title} {Compositional phase stability in medium-entropy and high-entropy {Cantor}-{Wu} alloys from an \textit{ab initio} all-electron {Landau}-type theory and atomistic modeling},}\ }\href {\doibase 10.1103/PhysRevB.105.115124} {\bibfield  {journal} {\bibinfo  {journal} {Physical Review B}\ }\textbf {\bibinfo {volume} {105}},\ \bibinfo {pages} {115124} (\bibinfo {year} {2022})}\BibitemShut {NoStop}%
\bibitem [{\citenamefont {Woodgate}(2024)}]{woodgate_modelling_2024}%
  \BibitemOpen
  \bibfield  {author} {\bibinfo {author} {\bibfnamefont {C.~D.}\ \bibnamefont {Woodgate}},\ }\href {https://doi.org/10.1007/978-3-031-62021-8} {\emph {\bibinfo {title} {Modelling {Atomic} {Arrangements} in {Multicomponent} {Alloys}: {A} {Perturbative}, {First}-{Principles}-{Based} {Approach}}}},\ \bibinfo {series} {Springer {Series} in {Materials} {Science}}, Vol.\ \bibinfo {volume} {346}\ (\bibinfo  {publisher} {Springer Nature Switzerland},\ \bibinfo {address} {Cham},\ \bibinfo {year} {2024})\BibitemShut {NoStop}%
\bibitem [{\citenamefont {Perdew}\ \emph {et~al.}(1996)\citenamefont {Perdew}, \citenamefont {Burke},\ and\ \citenamefont {Ernzerhof}}]{perdew_generalized_1996}%
  \BibitemOpen
  \bibfield  {author} {\bibinfo {author} {\bibfnamefont {J.~P.}\ \bibnamefont {Perdew}}, \bibinfo {author} {\bibfnamefont {K.}~\bibnamefont {Burke}}, \ and\ \bibinfo {author} {\bibfnamefont {M.}~\bibnamefont {Ernzerhof}},\ }\bibfield  {title} {\enquote {\bibinfo {title} {Generalized {Gradient} {Approximation} {Made} {Simple}},}\ }\href {\doibase 10.1103/PhysRevLett.77.3865} {\bibfield  {journal} {\bibinfo  {journal} {Physical Review Letters}\ }\textbf {\bibinfo {volume} {77}},\ \bibinfo {pages} {3865} (\bibinfo {year} {1996})}\BibitemShut {NoStop}%
\bibitem [{\citenamefont {Ebert}\ \emph {et~al.}(2011)\citenamefont {Ebert}, \citenamefont {Ködderitzsch},\ and\ \citenamefont {Minár}}]{ebert_calculating_2011}%
  \BibitemOpen
  \bibfield  {author} {\bibinfo {author} {\bibfnamefont {H.}~\bibnamefont {Ebert}}, \bibinfo {author} {\bibfnamefont {D.}~\bibnamefont {Ködderitzsch}}, \ and\ \bibinfo {author} {\bibfnamefont {J.}~\bibnamefont {Minár}},\ }\bibfield  {title} {\enquote {\bibinfo {title} {Calculating condensed matter properties using the {KKR}-{Green}'s function method—recent developments and applications},}\ }\href {\doibase 10.1088/0034-4885/74/9/096501} {\bibfield  {journal} {\bibinfo  {journal} {Reports on Progress in Physics}\ }\textbf {\bibinfo {volume} {74}},\ \bibinfo {pages} {096501} (\bibinfo {year} {2011})}\BibitemShut {NoStop}%
\bibitem [{\citenamefont {Soven}(1967)}]{soven_coherent-potential_1967}%
  \BibitemOpen
  \bibfield  {author} {\bibinfo {author} {\bibfnamefont {P.}~\bibnamefont {Soven}},\ }\bibfield  {title} {\enquote {\bibinfo {title} {Coherent-{Potential} {Model} of {Substitutional} {Disordered} {Alloys}},}\ }\href {\doibase 10.1103/PhysRev.156.809} {\bibfield  {journal} {\bibinfo  {journal} {Physical Review}\ }\textbf {\bibinfo {volume} {156}},\ \bibinfo {pages} {809} (\bibinfo {year} {1967})}\BibitemShut {NoStop}%
\bibitem [{\citenamefont {Matsumoto}\ and\ \citenamefont {Nishimura}(1998)}]{matsumoto_mersenne_1998}%
  \BibitemOpen
  \bibfield  {author} {\bibinfo {author} {\bibfnamefont {M.}~\bibnamefont {Matsumoto}}\ and\ \bibinfo {author} {\bibfnamefont {T.}~\bibnamefont {Nishimura}},\ }\bibfield  {title} {\enquote {\bibinfo {title} {Mersenne twister: a 623-dimensionally equidistributed uniform pseudo-random number generator},}\ }\href {\doibase 10.1145/272991.272995} {\bibfield  {journal} {\bibinfo  {journal} {ACM Transactions on Modeling and Computer Simulation}\ }\textbf {\bibinfo {volume} {8}},\ \bibinfo {pages} {3} (\bibinfo {year} {1998})}\BibitemShut {NoStop}%
\bibitem [{\citenamefont {Feller}(1968)}]{feller1968introduction}%
  \BibitemOpen
  \bibfield  {author} {\bibinfo {author} {\bibfnamefont {W.}~\bibnamefont {Feller}},\ }\href@noop {} {\emph {\bibinfo {title} {An Introduction to Probability Theory and Its Applications, Vol. 1}}}\ (\bibinfo  {publisher} {John Wiley \& Sons},\ \bibinfo {year} {1968})\BibitemShut {NoStop}%
\end{thebibliography}
\end{document}